\input harvmac
\newcount\figno
\figno=0
\def\fig#1#2#3{
\par\begingroup\parindent=0pt\leftskip=1cm\rightskip=1cm\parindent=0pt
\global\advance\figno by 1
\midinsert
\epsfxsize=#3
\centerline{\epsfbox{#2}}
\vskip 12pt
{\bf Fig. \the\figno:} #1\par
\endinsert\endgroup\par
}
\def\figlabel#1{\xdef#1{\the\figno}}
\def\encadremath#1{\vbox{\hrule\hbox{\vrule\kern8pt\vbox{\kern8pt
\hbox{$\displaystyle #1$}\kern8pt}
\kern8pt\vrule}\hrule}}

\overfullrule=0pt

%macros
%
\def\tilde{\widetilde}
\def\bar{\overline}
\def\Z{{\bf Z}}
\def\T{{\bf T}}
\def\S{{\bf S}}
\def\R{{\bf R}}

\font\zfont = cmss10 %scaled \magstep1

\def\bigone{\hbox{1\kern -.23em {\rm l}}}
\def\ZZ{\hbox{\zfont Z\kern-.4emZ}}

\def\CP{{\bf CP}}
\Title{hep-th/9610234, IASSNS-HEP-96-101}
{\vbox{\centerline{FIVE-BRANE EFFECTIVE ACTION}
\bigskip
\centerline{IN $M$-THEORY}}}
\smallskip
\centerline{Edward Witten\foot{Research supported in part
by NSF  Grant  PHY-9513835.}}
\smallskip
\centerline{\it School of Natural Sciences, Institute for Advanced Study}
\centerline{\it Olden Lane, Princeton, NJ 08540, USA}\bigskip

\medskip

\noindent

On the world-volume of an $M$-theory five-brane propagates a
two-form with self-dual field strength.  As this field is
non-Lagrangian, there is no obvious framework for determining
its partition function.  An analogous
problem exists in Type IIB superstring theory for the self-dual
five-form.  The resolution of these problems and definition of the
partition function is explained.  A more complete analysis of perturbative
anomaly cancellation for $M$-theory five-branes is also presented, 
uncovering some surprising details.
\Date{October, 1996}
%text of paper

\newsec{Introduction}

In a recent paper \ref\witten{E. Witten,
``Flux Quantization In $M$-Theory,'' hepth/9609122.}, 
it was shown that the low energy effective action
of $M$-theory on a closed eleven-dimensional spin manifold $Q$ is well-defined.
To be more precise, let $I_M$ be the Chern-Simons interaction
of the long wavelength limit of $M$-theory; schematically
$I_M=CGG+CI_8(R)$, where $C$ is the massless three-form, $G$ is the
gauge-invariant field strength of $C$, and $I_8(R)$ is a certain
eight-form constructed from the Riemann tensor (first obtained in
\ref\vw{C. Vafa and E. Witten, ``A One-Loop Test Of String Duality,'' Nuc.
Phys. {\bf B447} (1995) 261, hepth/9505053.}
and \ref\duff{M. J. Duff, J.-T. Liu, and R. Minasian,
``Eleven-Dimensional Origin Of String/String Duality: A One
Loop Test,'' hepth/9509084.}).  Let $\det D_{R.S.}$
be the path integral of the Rarita-Schwinger field\foot{As in the case
of any Majorana fermi field, this path integral might be more naturally
called a Pfaffian rather than a determinant.}; it is real but
not necessarily positive.  The long-wavelength limit of the quantum
measure of $M$-theory\foot{By the quantum measure I mean the
exponential of the effective action.}
 is a product of manifestly well-defined factors times
\eqn\normo{\det D_{R.S.} \,\,e^{iI_M},}
and it was shown in \witten\ that this product is well-defined (though
neither factor is well-defined separately).
The main novelty required for this result was a gravitational
shift in the quantization law of $G$.  The shifted quantization law says that
\eqn\umbo{\left[{G\over 2\pi}\right]={\lambda\over 2}+{\rm integral~cohomology
~class}}
where $[G/2\pi]$ is the cohomology class of $G/2\pi$, and $\lambda=p_1(X)/2$
($\lambda$ is integral for a spin-manifold $Q$).  

\nref\horava,{P. Horava and E. Witten, ``Heterotic And Type I
String Dynamics From Eleven Dimensions,'' hepth/9510209,
``Eleven-Dimensional Supergravity On A Manifold With Boundary,''
hepth/9603142.}
\nref\dealwis{S. De Alwis, ``Anomaly Cancellation In $M$-Theory,''
hepth/9609211.}
One would like to know whether the long-wavelength effective action
is still well-defined in the presence of impurities.  There are three
known kinds of impurities: boundaries of $Q$ (where $E_8$ supermultiplets
are believed to propagate); two-branes; and five-branes.  Anomaly
cancellation in the presence of boundaries has been demonstrated
in \refs{\horava,\dealwis}.\foot{To be more precise, 
perturbative anomaly cancellation  was analyzed in those papers; 
it would be natural to analyze the
global anomalies by an extension of ideas
in \ref\oldwitten{E. Witten, ``Global Gravitational Anomalies,''
Commun. Math. Phys. {\bf 100} (1985) 197, ``Global Anomalies
In String Theory,'' in {\it Anomalies, Geometry, and Topology},
ed. W. A. Bardeen and A. White (World Scientific, 1985).}.} 
Membrane world-volume anomalies are rather simple because
the world-volume theory is non-chiral.  The only issues concern the
sign of the path integral
of the world-volume fermions and the flux quantization
law for $G$.  There is potentially an anomaly affecting the sign of the 
fermion determinant, and it was shown in \witten\ that 
the $\lambda/2$ term 
in \umbo\ gives an additional effect that cancels this anomaly.

\nref\fivewitten{E. Witten, 
``Five-Branes And $M$-Theory On An Orbifold,'' hepth/9512219.}
\nref\town{P. Townsend, ``$D$-Branes From $M$-Branes,'' Phys. Lett.
{\bf B373} (1996) 68, hepth/9512062.}
\nref\strom{A. Strominger, ``Open $P$-Branes,'' Phys. Lett. {\bf B383}
(1996) 44, hepth/9512059.}
The remaining case,\foot{Still more generally, but beyond our scope
here, one could consider membranes ending on boundaries \horava\ or
 five-branes \refs{\town,\strom}.}
the five-brane, is the principal subject of the
present paper.  This case is particularly subtle because the five-brane
world-volume theory is chiral.  There are therefore potential perturbative
anomalies, which have been partially analyzed in \refs{\duff,\fivewitten}
and are considered more fully in sections three and five below.
One should also consider global anomalies, which can perhaps be treated
by an extension of the methods in \oldwitten.

But  a question more basic than those will be our main interest here:
{\it What is the five-brane partition function?}
This must be answered before the question of whether this partition
function has anomalies can be formulated.
The reason that there is a problem is simple.
One of the fields on the five-brane world-volume is a self-dual
three-form $T$; that is, on the world-volume there
is a two-form, say $\beta$, which
is constrained so that its field-strength $T$ is self-dual.  Such a field,
which we will call a chiral two-form,
has no Lagrangian formulation, and there is subtlety in defining,
even formally, what its partition function should be.

A chiral two-form in six dimensions poses a rather similar problem
to a chiral boson $\phi$ in two dimensions (whose field strength is a 
self-dual
one-form $\partial\phi$).  In fact, if the world-volume $W$ of the
five-brane is of the form $W=\Sigma\times {\bf CP}^2$,
\foot{This particular $W$ is 
not a spin manifold, but it could be embedded in an eleven-dimensional
spin manifold $Q$, and so can arise in $M$-theory.}   with $\Sigma$
a Riemann surface, then in the limit that $\Sigma$ is much larger
than the ${\bf CP}^2$, 
the theory of the chiral boson on $W$ reduces to the theory
of one chiral boson on $\Sigma$.   For one chiral boson, there is
no modular-invariant partition function -- a modular-invariant partition
function for chiral bosons requires an even self-dual lattice
\ref\goddard{P. Goddard and D. Olive, ``Algebras, Lattices,
and Strings,'' in {\it Vertex Operators In Mathematics And Physics,
Proceedings Of A Conference, November 10-17, 1983},''
eds. J. Lepowsky, S. Mandelstam, I. M. Singer (Springer-Verlag,
New York) p. 51.}, 
and in particular the net number of
chiral bosons must be a multiple of eight.  So this particular
example shows that it is impossible to define for the chiral two-form
a partition function that is ``modular-invariant,'' that is, invariant
under arbitrary diffeomorphisms of $W$.

We are not, however, dealing in $M$-theory
with a bare six-manifold $W$.  $W$ is embedded
in an eleven-dimensional spin-manifold $Q$, on which there are fields
obeying certain relations such as \umbo\ and \town\
$dT=G$.  In sections two and three, we demonstrate that this additional
data is exactly right to make it possible to define a partition function
for the chiral two-form, which is invariant (up to an ordinary anomaly)
not under all diffeomorphisms of $W$, but under those that preserve the
physical data.  (By ``an ordinary anomaly,'' we mean an anomaly that only
affects the phase of the partition function, and so can cancel anomalies
coming from other interactions or fields.)  

In section four, we make a digression from five-branes, and show that
a similar problem arises, and can be treated similarly, for the chiral
four-form of Type IIB superstrings -- that is, the massless field whose
field strength is a self-dual five-form.    This problem was avoided
in a previous analysis of Type IIB global anomalies \oldwitten\ by
considering only ten-manifolds of vanishing fifth Betti number; for
such space-times the problem does not arise, as will be apparent.

Finally, in section five, we look at the ordinary anomalies.
To be more precise, we give a more complete
analysis of five-brane perturbative anomalies than has been done
hitherto.  The analysis proves to require some surprising novelty,
and we actually get a complete answer only in the Type IIA case,
not in $M$-theory.

My interest in these issues 
 came originally from thinking about non-perturbative
superpotentials generated by five-brane instantons 
\ref\fiveinst{E. Witten,
``Non-Perturbative Superpotentials In String Theory,'' hepth/9604030}.  
Such an instanton contribution is roughly $e^{-\Phi} P$, where $\Phi$ is
a chiral superfield whose real part is the volume of the instanton
and $P$ comes from quantum fluctuations.  Roughly speaking,
zeroes of $P$ determine the supersymmetric vacua.  
In \fiveinst, the zeroes coming from fermion zero modes
were analyzed.  Additional zeroes can come from the behavior of
the partition function of the chiral two-form.  To determine the
locations of the zeroes requires the considerations of the present paper.

\newsec{Review Of Chiral Scalars In Two Dimensions}

\subsec{Basic Framework}

In this section, we review some aspects of
the theory of a chiral scalar in two dimensions, and explain
how some of the ideas generalize above two dimensions.

Since there is no effective Lagrangian formulation for a chiral scalar,
there is no natural way to use path integrals to determine the 
partition function.  One approach is to take a non-chiral
boson, which does have a Lagrangian and a well-defined partition function
$Z$, and try to write $Z$ as the absolute value squared of the chiral
boson partition function.  As is well known, things do not work so simply.
$Z$ is not the square of a holomorphic function but a sum of such squares;
the number of terms depends on the periodicity or radius
of the non-chiral boson.

As will become clear, the radius of relevance to our problem is the
free fermion radius.  This corresponds to momenta which take values in
the unimodular, but not even, one-dimensional 
lattice $\Z$, endowed with the quadratic
form $f(x)=x^2$ (that is, the unit element in $\Z$ is of length one).
At the free fermion radius, the partition function of the free boson
on a Riemann surface  $\Sigma$  of genus $g$ can be written
\foot{A thorough and direct study of this is in 
\ref\bost{L. Alvarez-Gaum\'e, J.-B.
Bost, G. Moore, P. Nelson, and C. Vafa, ``Bosonization On Higher Genus
Riemann Surfaces,'' Commun. Math. Phys. {\bf 112} (1987) 503.}.
Note that this problem is much simpler than the generic case
of rational conformal field theory in that the space of conformal
blocks has a distinguished basis, given by the $\Theta_\alpha$.}
\eqn\canwrit{Z=\sum_\alpha\left|\Theta_\alpha\right|^2.}
Here $\alpha$ runs over the spin structures on $\Sigma$, and
$\Theta_\alpha$ is the free fermion partition function with spin
structure $\alpha$.  In genus one, $\Theta_\alpha=\theta_\alpha/\eta$
where $\theta_\alpha$ is a theta function and $\eta $ is the Dedekind
eta function.
The functions $\Theta_\alpha$ are the candidate partition functions
of the chiral boson.  Our problem is to understand -- in a form suitable
for generalization to five-branes or to Type IIB in ten dimensions --
the fact that a choice of spin structure enables one to pick out a 
particular one of the $\Theta_\alpha$'s.

\def\S{{\bf S}}
\def\T{{\cal T}}
It is natural to couple the chiral
boson to a background gauge field $A$, which we can think of as a connection
on a line bundle $\T$.  The field strength is then
$\Lambda=d\phi+A$, and $\Lambda$ is no longer closed; it
obeys $d\Lambda=F$, with $F$ the field strength of $A$.  The existence
of a gauge-invariant field $\Lambda$ with $d\Lambda=F$ means
that (unless we make operator insertions that modify that equation)
$c_1(\T)$ must vanish, so $\T$ is topologically trivial.  Each choice
of connection $A$ gives (via the $\bar\partial$ operator) a complex
structure to $\T$.  The moduli space of such complex structures
is the Jacobian $J_\Sigma$ of $\Sigma$.  One can identify $J_\Sigma$ as
$H^1(\Sigma,\R)/H^1(\Sigma,\Z)$.

The coupling to a background gauge field
 has an analog in the five-brane problem.  The relevant background field here
is the three-form $C$ of eleven-dimensional supergravity (restricted
to the five-brane world-volume $W$).  
The field strength of $C$ is $G=dC$.  The self-dual three-form $T$ on $W$
 obeys \town\   $dT=G$, and this equation
is obviously quite analogous to $d\Lambda=F$.  The coupling to $C$ plays
for the self-dual three-form a role quite similar to the coupling
to a background gauge field for the self-dual one-form.  For reasons
that will be explained, we will be able to think of $C$ as defining a point
in the ``intermediate Jacobian'' $H^3(W,\R)/H^3(W,\Z)$.

The equation $dT=G$ means that the restriction of $G$ to the five-brane
world-volume $W$ must be zero cohomologically (just as above
the relation $d\Lambda=F$ implied that $\T$ is trivial topologically).
It follows, therefore,
given \umbo, that {\it the restriction of $\lambda$ to $W$ is even}.
This will be important in section three.

\def\L{{\cal L}}
\bigskip\noindent
{\it Embedding In Non-Chiral Theory}

We will look now more closely at the embedding of the chiral scalar
in the theory of a non-chiral scalar.    In doing so, we work on a 
two-dimensional surface $\Sigma$ of {\it Euclidean} signature, so
the Lagrangian will be complex, and the self-duality condition 
reads $d\phi =i*d\phi$.  

We consider thus a scalar field $\phi$, with a periodicity $\phi\to\phi+2\pi$,
and a Lagrangian (at the free fermion radius) 
\eqn\normo{L={1\over 8\pi}\int d^2x\sqrt g g^{ij}\partial_i\phi\partial_j\phi.}
We introduce as explained above
a $U(1)$ gauge field $A$ with gauge transformation law
$\delta A_i=-\partial_ia$, $\delta \phi=a$, and Lagrangian\foot{A more
complete account of the point of view that follows, in the more general
context of the WZW model, is in \ref\facto{E. Witten, ``On Holomorphic
Factorization Of The WZW Model,'' Commun. Math. Phys. {\bf 144} (1992) 189.}.} 
\eqn\gormo{L={1\over 8\pi}\int_\Sigma d^2x\sqrt g g^{ij}(\partial_i\phi+A_i)
(\partial_j\phi+A_j) +{i\over 4\pi}\int_\Sigma \phi\epsilon^{ij}\partial_iA_j.}
(The transformation law of $\phi$ means that $\phi$ is a section of the
circle bundle on which $A$ is a connection, which therefore must be
topologically trivial as explained earlier in a somewhat different way.)
The point of this particular coupling is that $A$ has been coupled only
to the chiral part of $\phi$. This can be made clear by introducing
local complex coordinates $z,\bar z$ (with orientations so that
$\epsilon_{z\bar z}=-\epsilon_{\bar z z}=i$)
and expanding the Lagrangian to get
\eqn\pormo{L={1\over 4\pi} \int_\Sigma |dz\wedge d\bar z| 
\left(\partial_z\phi\partial
_{\bar z}\phi +2\partial_z\phi A_{\bar z} +A_zA_{\bar z}\right).}
Thus, only $A_{\bar z}$ and not $A_z$ couples to the quantum field $\phi$.
It follows that in a suitable sense the partition function
\eqn\ohormo{Z(A)=\int D\phi \,e^{-L}}
depends  holomorphically on $A_{\bar z}$.   

To be more precise, introduce
a complex structure on the space of gauge fields in which $A_{\bar z}$
is holomorphic and $A_z$ is anti-holomorphic.   
A holomorphic line bundle $\L$ over the space of gauge fields can be 
defined by taking $\L$ to be the trivial line bundle endowed with
the covariant derivatives
\eqn\hungo{{D\over DA_i}={\delta\over\delta A_i}+
{i\epsilon_{ij}\over 4\pi}
A_j.}  
This means that 
\eqn\iponggo{{D\over DA_z}={\delta\over\delta A_z}+
{A_{\bar z}\over 4\pi}.} 
To show that this connection defines a {\it holomorphic} 
structure on $\L$, one must check from \hungo\
that 
\eqn\hh{\left[{D\over DA_z(x)},{D\over DA_z(x')}\right]=0.}
This is straightforward.
Then using the fact 
that $A_z$ appears in $L$ only in the $A_zA_{\bar z}$ term
in \pormo, together with the explicit form of \iponggo,
one finds that $(D/DA_z)e^{-L}=0$.  The
partition function $Z=\int D\phi\,e^{-L}$
therefore obeys the same equation:
\eqn\obungo{{D\over DA_z} Z=0.}
Thus, {\it the partition function is a holomorphic section of $\L$,} 
over the space of all connections.

The fact that the partition function is most naturally seen as a 
section of a line bundle rather than a function is related to the fact
that the Lagrangian \gormo\ is not gauge-invariant.  Under gauge
transformations $\delta\phi=a$, $\delta A=-da$, one has
\eqn\duggo{\delta L={i\over 4\pi}\int_\Sigma a\,\epsilon^{ij}\partial_iA_j.}
The partition function thus obeys not standard gauge-invariance, which
would read
\eqn\orgo{\partial_i{\delta\over \delta A_i}Z=0,}
but rather 
\eqn\pporgo{\left(
\partial_i{D\over DA_i} -{i\epsilon^{ij}F_{ij}\over 4\pi}\right)Z=
0.}
This means that the partition function $Z$ is invariant under
infinitesimal gauge transformations if interpreted as a section
of $\L$ rather than as a function.  The operators on the left
hand side of \pporgo\ are the ones that generate infinitesimal
gauge transformations when acting on sections of $\L$.  This is a special
case of a more general assertion about the WZW model; see \facto, eqn. (2.17).

So far we have coupled to arbitrary (topologically trivial) background
gauge fields $A$.  The partition function is constrained by the two
conditions of gauge covariance \pporgo\ and holomorphy \obungo.
Taken together, the two conditions determine how the partition
function transforms under a {\it complex} gauge transformation
$\delta A_{\bar z}=-\bar\partial \epsilon,$ $\delta A_z=-\partial \bar \epsilon
$.  By a complex gauge transformation one can reduce to $F=0$.  So
there is no loss of generality in considering only the coupling to
background fields with $F=0$. 

The gauge field $A$, modulo infinitesimal gauge transformations and
with $F=0$, defines a point in $H^1(\Sigma,\R)$.  
With infinitesimal gauge transformations acting as in \pporgo,
$\L$ descends to a line bundle -- which we will also call
$\L$ -- on $H^1(\Sigma,\R)$.  $Z$, being gauge invariant in the sense
of \pporgo, descends to a section of $\L$ over $H^1(\Sigma,\R)$.
However, we want to also divide by the ``big gauge transformations,''
and interpret $Z$ as a section of a line bundle over $H^1(\Sigma,\R)/
H^1(\Sigma,\Z)$, which is the Jacobian $J_\Sigma$ of $\Sigma$.

There is no natural choice of how the ``big gauge transformations''
should act on $\L$.  Why this is so is explained below.
There are in fact different and equally natural
line bundles $\L_\alpha$, obtained by differing choices of
how the big gauge transformations act on $\L$.

Related to this, the partition function $Z$ considered 
so far is not really the object we want.  It includes
the contributions of the ``wrong chirality'' part of $\phi$ which
though decoupled from $A$ is still present in the partition function.
If one tries to carry out holomorphic factorization to suppress the
wrong chirality field, one finds as in \canwrit\ that
$Z$ is a sum of terms, each involving (as will become clear)
a different $\L_\alpha$.

What has been
said so far applies to a chiral $2k$-form $\beta$ in $4k+2$ dimensions for any
$k$.  The arguments have been presented in such a way that they
carry over without any essential change.  We repeat the story briefly.
Letting $C$ be a background $2k+1$ form on a $4k+2$-dimensional manifold $W$,
and $G=dC$,  consider
a non-chiral $2k$-form $\beta$ on $W$, with coupling to $C$ given schematically
by $L_\beta=\int_W(|d\beta+C|^2+\beta\wedge G)$.  Choose
the  coefficients in $L_\beta$ chosen so that
only the anti-self-dual part of $C$ couples.  The ``wrong chirality'' part
of $\beta$ is thus present, but decoupled.  Under infinitesimal gauge
transformations $\delta \beta = \alpha$, $\delta C=-d\alpha$, (with
$\alpha$ a $2k$-form), $L_\beta $ changes by  
\eqn\onorko{\delta L_\beta \sim \int_W \alpha\wedge G.}
This means that the partition function $Z$ should be understood as a section
of a line bundle ${\cal L}$ over the space of $C$'s.  The fact that only
the anti-self-dual part of $C$ couples means that 
$Z$ is actually a holomorphic section.  Holomorphy plus gauge-invariance
imply invariance under complexified gauge transformations, which can
be used to reduce to the case that $C$ is a harmonic $2k+1$-form,
defining a point in $H^{2k+1}(W,\R)$.  The partition function is thus
naturally induced from a section of a line bundle over $H^{2k+1}(W,\R)$.  To
define the theory of the chiral two-form, one must carry out holomorphic
factorization, throw away the anti-chiral contribution, and divide by 
``big gauge transformations'' so as to descend to  $J_W=H^{2k+1}(W,\R)/
H^{2k+1}(W,\Z)$, 
which is known as the ``intermediate Jacobian'' of $W$.  Holomorphic
factorization leads to a sum of different terms, each associated with
a different line bundle on $J_W$.  The key to finding the partition function
of the self-dual scalar is to find a way to pick out a particular
term from this sum, or equivalently a particular line bundle on $J_W$.  

\bigskip\noindent
{\it The Line Bundle}

Now let us go back to the free fermion approach to the non-chiral scalar
in two dimensions.
The factorization \canwrit\ still holds after coupling to $A$, but now
the $\Theta_\alpha$ are functions of $A$ as well as of the complex structure
$\tau$ of $\Sigma$.\foot{If $A$ is coupled chirally, as above, the
$\bar\Theta_\alpha$ still depend on $\tau$ only; if one uses a vector-like
coupling of $A$, the $\bar\Theta_\alpha$ would still be the complex
conjugates of the $\Theta_\alpha$.}   

The $\Theta_\alpha$ can be written as
$\Theta_\alpha(\tau,A)=\theta_\alpha(\tau,A)/\tilde \eta(\tau)$
where the $\theta_\alpha(\tau,A)$ are theta functions on the Jacobian 
$J_\Sigma$
(we recall that each $A$ determines a point on $J_\Sigma$)
and  $\tilde\eta(\tau)$ depends only on $\tau$.

Because the coupling of a chiral scalar to a gauge field
violates gauge invariance,   the partition functions $\Theta_\alpha(\tau,A)$,
in their dependence on  $A$, are not naturally understood as functions
but as sections of appropriate line bundles ${\cal L}_\alpha$ over
the Jacobian.  We have essentially seen this already from the
bosonic point of view.
Moreover -- as already indicated -- the $\theta_\alpha$
of different $\alpha$ are all sections of {\it different} line 
bundles over $J_\Sigma$.  That is the key to our problem.  It means that once
one finds the line bundle ${\cal L}_\alpha$ on the Jacobian, 
a corresponding partition function $\Theta_\alpha$ of the chiral scalar
is naturally determined.  In fact, each ${\cal L}_\alpha$ has (up to
a complex multiple) only one holomorphic section, as we will see, 
and so automatically determines its own theta function $\Theta_\alpha$.
\foot{It may appear that  the uniqueness (up to an $A$-independent
but possibly $\tau$-dependent multiple) of the holomorphic section of 
${\cal L}_\alpha$ determines only the $A$ dependence and not the $\tau$
dependence of the chiral partition function $\Theta_\alpha$. 
But the partition function $\Theta_\alpha$
obeys a heat equation (a special case
of the KZB equation obeyed by the conformal blocks of the WZW model)
that determines its $\tau$ dependence when the $A$ dependence is known.
The heat equation is a consequence of the Sugawara construction:
the stress tensor of the chiral boson is the square of the current
$\partial_z\phi+A_z$ that couples to $A_{\bar z}$.  Likewise,
the stress tensor of the chiral
$2k$-form is a quadratic expression in the current $(d\beta+C)^+$ 
(the superscript $+$ refers to a projection on the self-dual part),
as a result of which there is  
a heat equation that determines the  dependence on the metric of
$W$ from
the $C$ dependence.   }
This approach, in which the partition function is defined by first
finding the right line bundle,
may at first sound esoteric, but can be implemented
quite uniformly for our three cases: the chiral scalar in two dimensions;
the self-dual three-form on the five-brane world-volume; and the self-dual
five-form of Type IIB theory in ten dimensions.

\subsec{ Line Bundles On The Jacobian}

Our problem, then, is to study line bundles on the Jacobian $J_\Sigma$
of a Riemann surface $\Sigma$, or the intermediate Jacobian $J_W$ of
a  $4k+2$-dimensional manifold $W$.

To begin with, consider more generally
a $2n$-dimensional torus $J=\R^{2n}/\Gamma$,
where $\Gamma$ is a rank $2n$ lattice in $\R^{2n}$.  A ``principal
polarization'' of $J$ is an element $\omega\in H^2(J,\Z)$ such that
\eqn\jurry{\int_J{\omega^n\over n!}=1.}
$\omega$ can be represented by a two-form on $J$ which is uniquely 
determined if we require it to be invariant under translations of $J$;
this will always be assumed.

An example of such an $\omega$ is as follows.  Let $x^i,y_j$, $i,j=1,\dots, n$
be coordinates on $\R^{2n}$ such that $\Gamma$ is spanned by unit vectors
$e_i$ and $f^j$ in the $x^i$ and $y_j$ directions, respectively.
Then $\omega=\sum_idx^i\wedge dy_i$ defines a principal polarization.
Conversely, any translation-invariant two-form $\omega$ representing
a principal polarization can be put in such a form by a suitable choice
of coordinates.

In the example just given, the pairings of the two-form $\omega$ with
the vectors $e_i$ and $f^j$ are
\eqn\ompo{\omega(e_i,f^j)=\delta_i{}^j,\,\,\omega(e_i,e_j)=\omega(f^i,f^j)
=0.} 
Thus, on the lattice $\Gamma$, $\omega$ defines an integer-valued antisymmetric
pairing which is non-degenerate and ``minimal'' (the Pfaffian of $\omega$ is
as small as possible -- equivalent to the assertion \jurry\ that the volume
of a unit cell is 1).  

If $J$ is the intermediate Jacobian of a $4k+2$-dimensional manifold 
$W$ (so $J=H^{2k+1}(W,\R)/H^{2k+1}(W,\Z)$), then $\Gamma$ corresponds
to the lattice $H^{2k+1}(W,\Z)$ (mod torsion).  The intersection
pairing or cup product $H^{2k+1}(W,\Z)\times H^{2k+1}(W,\Z)
\to H^{4k+2}(W,\Z)\cong \Z$ defines an integer-valued antisymmetric
pairing on $\Gamma$ which is isomorphic to \ompo\ in suitable coordinates.
(For instance, if $W$ is a Riemann surface, then the intersection pairing
on $H^1$ can be put in the form \ompo\ by a choice of $A$ and $B$ cycles.)
Therefore, the intermediate Jacobians of interest to us 
are always  naturally endowed with a principal polarization.
The curvature form $\omega$ associated with this polarization actually
can be seen in the curvature of the connection introduced in \hungo.

Given a metric on $W$, the intermediate Jacobian $J_W$ has a natural
metric defined as follows.  
The tangent space to $J_W$ is the space ${\cal W}$ 
of harmonic $2k+1$-forms
on $W$.  The Hodge $*$ operator maps ${\cal W}\to {\cal W}$ 
with $*^2=-1$, so it defines a complex structure on $J_W$.
A metric on $J_W$ can be defined by saying that for $C\in {\cal W}$,
$||C||^2=\int_W C\wedge *C$.
This metric on $J_W$ is translation-invariant and Kahler.
The associated Kahler form is our friend the polarization of $J_W$,
$\omega(C,C')=\int_WC\wedge C'$.  In particular, $\omega$ is
of type $(1,1)$ and positive in this Kahler metric.  

In fact, the partition function of the chiral $2k$-form only depends
on the metric  on $J_W$ and the line bundle $\L$, up to elementary
factors determined by the anomalous Ward identities.  This is a striking
simplification, as the metric on $W$ (on which the partition function
might depend {\it a priori}) depends on infinitely many parameters, but
the metric on $J_W$, being translation-invariant, depends on only
finitely many parameters.  This result 
can be seen using the heat equation described
in the footnote at the end of section 2.1: since the stress tensor
is quadratic in the ``currents,'' the response to a change in the 
metric $W$ can be expressed in terms of the response to a change in the
background $C$-field, and just as for a chiral scalar in two dimensions,
the only non-elementary terms that arise are those that involve the change in
metric of $J_W$ under change in metric of $W$.

\bigskip\noindent
{\it From Polarization To Line Bundle}

Since line bundles are classified topologically by their first Chern
class, there is topologically
up to isomorphism a unique line bundle $\L$ on $J_W$ whose
first Chern class is $c_1(\L)=\omega$.  However, we need to describe
$\L$ much more precisely.  We want to find a $U(1)$ connection $B$ on $\L$,
whose curvature $F=dB$ equals $2\pi\omega$.  
This would lead as follows to a definition of the partition function of
the chiral $2k$-form.	Since $\omega$ is of type
$(1,1)$ in the complex structure on $J_W$, 
the connection $B$ determines a complex structure on $\L$.
The index of the $\bar\partial$ operator on $J$, with values in $\L$, is
\eqn\indbar{\sum_{i=0}^{\dim_{\bf C}\,J_W}(-1)^i\dim H^i(J_W,\L)
=\int_Je^{c_1(\L)}{\rm Td}(J_W)=\int_Je^{\omega}=1.}
 (Here Td is the Todd genus; since $J_W$ has a flat
metric, ${\rm Td}(J_W)=1$.  We also use \jurry.)  Since $\omega$ is positive, 
the cohomology $H^i(J_W,\L)=0$ for $i>0$,\foot{This is
true by the Kodaira vanishing theorem,
which uses the fact that $\bar\partial^*\bar\partial+
\bar\partial\,\bar\partial^*$ is strictly positive, in this situation, 
for $i>0$.}  so the index formula actually asserts that $H^0(J_W,\L)$ is
one-dimensional.  Thus, the line bundle $\L$ has (up to a complex multiple)
a single holomorphic section.  This section is the desired partition function
of the chiral $2k$-form, at least as regards the $C$ dependence.
But since the different terms in the holomorphic factorization of the
non-chiral $2k$-form have different $C$ dependence, once we know
the $C$ dependence we are essentially done.  (More fundamentally, as
remarked in a previous footnote, the $\tau$-dependence can be determined
once the $C$-dependence is known from the fact that the chiral partition
function obeys an analog of the KZB equation for conformal blocks of
the WZW model.) 

So our basic problem is really to find a $U(1)$ gauge field on $J_W$ whose
curvature is $F=2\pi\omega$.   Now, on a simply-connected manifold,
a $U(1)$ gauge field $B$ is determined up to isomorphism by its curvature.
We are dealing instead with the torus $J=\R^{2n}/\Gamma$.  
(In what follows, $J$ can be any torus with a principal polarization,
as opposed to $J_W$ which has additional structure such as a metric and
complex structure.   So  we drop the subscript $W$ for the time being.) 
To fix $B$ we must give, in addition to the curvature, the holonomies around
noncontractible cycles in $J$.  

If $a$ is a lattice point in $\Gamma\subset \R^{2n}$, then
 the straight line from the origin in $\R^{2n}$ to $a$
determines a closed curve $C(a)$ in $J$.  Let $H(a)=\exp(i\int_{C(a)}B)$
be the holonomy of $B$ around $C(a)$.  $B$ will be completely fixed
if the $H(a)$ are given.  We would like to pick the $H(a)$'s to preserve,
as much as possible, the invariance under the symplectic group
${ Sp}(2n,\Z)$ (which acts on $\Gamma$ preserving $\omega$).
The most obvious choice would be $H(a)=1$ for all $a$.  This
is impossible for the following reason.

The $H(a)$'s are constrained as follows.
If $a$ and $b$ are any two lattice points, then
\eqn\keycon{H(a+b)=H(a)H(b)(-1)^{\omega(a,b)}.}
This is obtained as follows.  The lattice points $0,a,b$, and $a+b$
are vertices of a parallelogram $\Delta_{a,b}$
through which the magnetic flux
is $2\pi \omega(a,b)$.  
The lattice points $0,a$, and $a+b$ are vertices
of a triangle $T_{a,b}$ which 
is just half of $\Delta_{a,b}$; the magnetic flux through $T_{a,b}$ is
$\pi\omega(a,b)$.  The sides of $T_{a,b}$ are $C_a$, $C_b$, and $C_{a+b}$.
So \keycon\ is the usual relation, following from Stokes's theorem, between
the magnetic flux through a surface $S$ -- in this case $T_{a,b}$ --
and the holonomy around the boundary
of $S$.

\def\M{{\cal M}} A basic question is now whether it is possible
to pick a line bundle in a way invariant under the $Sp(2n;\Z)$ that
acts on $J$ preserving the polarization.
\keycon\ shows that this is impossible; 
$Sp(2n;\Z)$ would require that $H(a)=H(b)$ for all primitive
lattice vectors $a$ and $b$, and this is incompatible with \keycon.
\foot{There is one significant exception; if $J$ is two-dimensional
then \keycon\ allows $H(a)=-1$ for all primitive $a$.  This corresponds
to the fact that on (and only on) a genus one curve, there 
is a spin structure that is completely diffeormorphism-invariant, namely
the ``odd'' one (the trivial spin bundle).  For $\dim\,J>2$, the 
possibility $H(a)=-1$ for all primitive $a$ is
excluded as one can find vectors $a,b$ with $a,b$, and $a+b$ all 
primitive and $\omega(a,b)=0$.}
In particular, \keycon\ does not permit us to take $H(a)=1$ for all $a$.
\keycon\ does, however, allow $H(a)^2=1$ for all $a$.  The numbers
$H(a)^2$ would be the holonomies around $C_a$
of the connection $2B$ on the line bundle
$\L^{ 2}$, which we will call ${\cal M}$.  
Thus, there is a completely natural, 
$Sp(2n,\Z)$-invariant line bundle
${\cal M}$ with holonomy $+1$ around each $C_a$ and first Chern class
$2\omega$.  The factor of 2 means that (using the index formula as above)
$H^0(J,\M)$ is of dimension $2^g$.  
We need a line bundle $\L$ of first Chern
class $1\cdot \omega$, with just one holomorphic section, which will
be our partition function. 

In searching for an $\L$ that 
is ``as canonical as possible,'' we can require
that $\L^2$ is isomorphic to $\M$, or equivalently that the holonomies
$H(a)$ of $\L$ are all $\pm 1$.  The number of such $\L$'s is $2^{2n}$;
they differ by $H(a)\to H(a)(-1)^{\epsilon(a)}$ with $\epsilon\in 
H^1(J,\Z_2)$.  Thus, the search for $\L$ is reduced to the selection
among a finite set of possibilities.

In case $J=J_\Sigma$ is the Jacobian of a Riemann surface $\Sigma$,
we already know from bose-fermi equivalence (that is, from the holomorphic
factorization of the free boson at the free-fermion radius)
that to pick an $\L$ out of the $2^{2n}$ possibilities, what we need
is precisely a spin structure on $\Sigma$.  In the remainder of this
section, I will sketch three direct explanations of this fact.
The first two, though not needed in the rest of the paper (and therefore
not explained below in much detail), are included
because they are short and illuminating.  The third explanation,
though not new \ref\dijk{R. Dijkgraaf and E. Witten, ``Topological
Gauge Theories And Group Cohomology,'' Commun. Math. Phys. {\bf 129} 
(1990) 393.}, is perhaps
less well-known.  It is this third approach that we will later generalize
above two dimensions.

\bigskip\noindent
{\it The Determinant Bundle}

The first approach is
 directly related to the free fermion construction of the chiral boson.
If we are given a spin structure $\alpha$ on $\Sigma$,
then we can use the determinant of the Dirac operator to obtain a 
line bundle on $J_\Sigma$.  

Thus, given a flat connection $A$ on $\Sigma$ representing a point in
$J_\Sigma$, let $\T_A$ be the corresponding
flat line bundle on $\Sigma$.  Let $D_\alpha(A)$ be the Dirac operator with
values in $\T_A$, using the spin structure $\alpha$. 
The determinant line of $D_\alpha(A)$ is a complex
line $\L_{\alpha,A}$, and these fit together as $A$ varies to give the desired
line bundle $\L_\alpha\to J_\Sigma$, which can be shown via
index theory to have the right first Chern class.
\foot{An important subtlety, which is the 
reason that one must here use the Dirac operator rather than the 
$\bar\partial$ operator (which does not require a choice of spin structure)
is that because $D_\alpha(\T_A)$ 
has zero index, $\L_A$ depends in an appropriate
sense only on the isomorphism class of $\T_A$.  That is essential because
without making any arbitrary choices, $J_\Sigma$ parametrizes a family
of isomorphism classes of line bundles on $\Sigma$, but 
non-modular-invariant choices are needed to get an actual family of line
bundles.}  It thus has a single holomorphic section, which is in
fact the function $\Theta_\alpha$ that appears in holomorphic factorization
of the non-chiral boson.

So in particular, a choice of spin structure on $\Sigma$
gives a choice of line bundle on $J_\Sigma$.  Of course, the discussion
has now brought us back to our starting point \canwrit, and if we were
interested only in the two-dimensional case we could have spared much
of our effort.

\def\SS{{\cal S}}
\bigskip\noindent
{\it The Shifted Jacobian}

Now we consider briefly
another approach which  is less obviously related to physics.
We consider the shifted Jacobian $J_{\Sigma,n}$
of $\Sigma$, which parametrizes holomorphic line bundles on $\Sigma$
of degree $n$.
They are all non-canonically isomorphic to the ordinary Jacobian $J_\Sigma$.
Fixing a line bundle $\SS$ of degree $n$, the map $\T\to \T\otimes \SS$
(where $\T$ is a line bundle of degree zero, defining a point in $J_\Sigma$, 
and
$\T\otimes \SS$ therefore has degree $n$ and defines a point in $J_{\Sigma,n}$)
is an isomorphism between $J_\Sigma$ and $J_{\Sigma,n}$.  In particular,
the existence of this isomorphism means that each $J_{\Sigma,n}$ is
naturally endowed with a principal polarization $\omega$.

On $J_{\Sigma,g-1}$, there is actually a completely natural 
(``modular-invariant'') choice of
line bundle  with first Chern class $\omega$; we call it $\L'$ to
distinguish it from the desired line bundle $\L$ on $\Sigma$.
In fact, a line bundle on a complex manifold can be given by specifying
a divisor.  On $J_{\Sigma,g-1}$, there is a natural divisor, the $\Theta$
divisor, that parametrizes line bundles with a holomorphic section.  It
can be shown that the associated line bundle $\L'$ on $J_{\Sigma,g-1}$ 
has first Chern class $\omega$.

Now if $\SS$ is {\it any} line bundle on $\Sigma$ of degree $g-1$, then
by using $\SS$ as explained above
to establish an isomorphism  between $J_\Sigma$ and $J_{\Sigma,
g-1}$, we can interpret $\L'$ as a line bundle
on $J_\Sigma$.  Thus, to find the desired line bundle on $J_\Sigma$
all we need to do is to pick an $\SS$.  There is no natural  
choice of $\SS$.  The closest one can come is to set 
$\SS$ equal to one of the $2^{2g}$ spin structures of $\Sigma$.
(Recall that a spin structure on $\Sigma$ corresponds to a line
bundle of degree $g-1$ whose square is isomorphic to the canonical
line bundle; there are $2^{2g}$ of them, and they are on the smallest
possible orbit of the modular group in $J_{\Sigma,g-1}$.)  So we get
again the expected result: any choice of spin structure gives 
a choice of line bundle $\L$ on $J_\Sigma$ with first Chern class $\omega$.

\bigskip\noindent{\it Chern-Simons Theory}

Finally, we come to the approach that we will actually use in the
rest of this paper, in generalizing above two dimensions.  To construct
the desired line bundle on the Jacobian of a {\it two}-dimensional surface,
we use the Chern-Simons functional of a gauge field in {\it three} dimensions.

Let $M$ be a closed oriented three-manifold, and let $A$ be a connection on
a $U(1)$ bundle $\T$ over $M$.  If $\T$ is topologically trivial,
so that in a given gauge $A$ is an ordinary one-form, the Chern-Simons
functional is
\eqn\nomore{I(A)={1\over 2\pi}\int_M\epsilon^{ijk}A_i\partial_jA_k.}
If $\T$ is topologically non-trivial, a more powerful approach to
defining $I$ is needed.  
Let $X$ be an oriented four-manifold
with boundary $M$, over which $A$ and $\T$ extend, and pick such an extension.
\foot{Since line bundles are classified by maps to $\CP^{\infty}$,
the existence of such an $X$ follows from the statement
that the oriented
bordism group $\Omega_3(\CP^\infty)$ vanishes.  In fact,
a more precise statement (which we will need presently) also holds:
the spin bordism group $\Omega_3(\CP^\infty)$ vanishes.
(This means that if $M$ is a spin manifold with a given spin structure,
one can choose $X$ with boundary $M$
so that $A$ and the spin structure of $M$ extends over $X$.)
The following proof of this was sketched by P. Landweber.  According  
to the proposition on p. 354 of \ref\stong{R. E. Stong, {\it Notes
On Cobordism Theory} (Princeton University Press, 1968).}, 
$\Omega_3(\CP^\infty)=\Omega_5^{Spin^c}$.  According to the
theorem on p. 337 of the same book, since there are no rational
characteristic numbers in odd dimensions, 
$\Omega_5^{Spin^c}$ is determined by Stieffel-Whitney
numbers.  A $Spin^c$ manifold has $w_1=w_3=0$.  A five-dimensional
manifold with $w_1=0$ has $w_5=0$.  So a five-dimensional $Spin^c$
manifold has no non-zero Stieffel-Whitney numbers, and hence
$\Omega_5^{Spin^c}=0$.} (We assume that the orientation of $X$ is related
to that of $M$ by a definite convention, for instance ``outward 
normal first.''  We will abbreviate the statement that $X$ has boundary
$M$ and $(A,\T)$ have been extended over $X$ by saying that
$X$ has boundary $(M,A)$.)  Then define
\eqn\omore{I_X(A)={1\over 2\pi}\int_X\epsilon^{ijkl}\partial_iA_j\partial_kA_l
={1\over 8\pi}\int_X\epsilon^{ijkl}F_{ij}F_{kl}.}
The point of this definition is, first of all, that if $\T$ is trivial
and $A$ is well-defined as a one-form, then by Stokes's theorem,
$I_X(A)$ coincides with $I(A)$ as defined in \nomore.  Furthermore,
$I_X(A)$ is defined even if $\T$ is topologically non-trivial.  
What remains is to investigate the extent to which $I_X(A)$ depends
on the choice of $X$ (and the extension of $A$ and $\T$).  Given another
oriented four-manifold $X'$ with boundary $(M,A)$, 
one can glue $X$ and $X'$ together along their
common boundary to make a closed four-manifold $Y$ with
a $U(1)$ gauge field $A$; if we reverse
the orientation of $X'$, then the orientations of $X$ and $X'$ match along
their common boundary, so that $Y$ has a natural orientation.
Then $I_X(A)-I_{X'}(A)=I_Y(A)$, where
\eqn\tomore{I_Y(A) = {1\over 8\pi}\int_Y\epsilon^{ijkl}F_{ij}F_{kl}.}
The point is now that because the cohomology class $[F/2\pi]$ is
integral, $I_Y(A)=2\pi\cdot {\rm integer}$.  Hence $I_X(A)$ is independent
of $X$ modulo $2\pi\Z$; with the understanding that there is this
$2\pi\Z$ ambiguity, we henceforth drop the subscript $X$ and refer simply
to $I(A)$.

The fact that $I(A)$ is well-defined modulo $2\pi$   means that 
\eqn\trufo{ e^{iI(A)} }
is well-defined.  $U(1)$ Chern-Simons gauge theory at level one is in
fact defined by the path integral
\eqn\rufo{\int DA\,\,e^{iI(A)}.}

If one considers $M$ to be not a closed three-manifold, but rather
$M=\Sigma\times \R$ where $\Sigma$ is a Riemann surface and $\R$
parametrizes ``time,'' then the moduli space of classical solutions
of the Chern-Simons theory is the Jacobian $J_\Sigma$ (since
the equation for a critical point of $I(A)$ is $F=0$).  The quantum
Hilbert space is a space of sections of a certain line bundle $\M$ over
$J_\Sigma$.  $\M$ is roughly the sort of object that we are looking for.
However, since the construction is completely diffeomorphism-invariant
and in particular no choice of spin structure has entered,
 $\M$ is  modular-invariant and therefore cannot be
the desired line bundle with first Chern class $\omega$.  $\M$ is
in fact, as we will see, 
the modular-invariant line bundle with first Chern class
$2\omega$ (and all $H(a)=1$) that was seen before.

To better understand  $\M$, a first orientation is as follows.
Let $\Sigma$ be a Riemann surface and let $N$ be a three-manifold
with boundary $\Sigma$.  Let $A$ be a connection on a line bundle $\T$ over
$N$, and consider the (level one) Chern-Simons Lagrangian
\eqn\omibo{L_{C.S.}=-iI(A)=-{i\over 2\pi}\int_N\epsilon^{ijk}A_i\partial_jA_k.}
In proving gauge-invariance, one must integrate by parts, and one picks
up a 
non-zero surface term because $N$ has a non-empty boundary.  In fact, under
$\delta A_i=-\partial_ia$, one finds 
\eqn\bimo{\delta L_{C.S.} = 
{i\over 2\pi}\int_\Sigma a \epsilon^{ij}\partial_iA_j.}
Just as in our discussion of the chiral boson, this violation of
gauge invariance means that $e^{-L_{C.S.}}=e^{iI(A)}$ is most naturally
understood not as a function but as a section of a line bundle $\M$ over
the space of gauge fields on $\Sigma$. 
In fact, the right hand side of \bimo\ has the same form as \duggo, but
the coefficient is twice as large; the factor of two means that $\M$
is not the desired line bundle $\L$ for the theory of the chiral boson,
but rather $\M=\L^2$, as we will show more fully later.

Let us describe more precisely the  construction of $\M$.  
First I sketch a rather down-to-earth approach.  To describe up to isomorphism
a line bundle $\M$, with $U(1)$ connection, on any given
manifold $Z$, it suffices
to define the holonomies of the connection around an arbitrary loop in $Z$;
these must obey certain axioms that will be discussed.
In our case, $Z$ is the space of $U(1)$ gauge fields on $\Sigma$.
Suppose we are given a loop $C$ in the space of gauge fields, that is to say
a family of gauge fields on $\Sigma$  depending on an extra parameter $\theta$
($0\leq\theta\leq 2\pi$); $\theta$ parametrizes the position on $C$.  Making
the $\theta$ dependence explicit, we write $A_i(x;\theta)$ for this
family of gauge fields,      where $x$ is a point in $\Sigma$.  Now
on the three-manifold $\Sigma\times \S^1$, we introduce the gauge
field $ A_C$ whose component in the $\S^1$ (or $\theta$) direction
is zero, and whose components along $\Sigma$ are $A_i(x;\theta)$.
We define the holonomy of $\M$ around the loop $C$ to be
\eqn\hofc{H(C)=e^{iI( A_C)}.}
The property of the $H(C)$'s that is needed for them to be the holonomies
of  a connection on a line bundle $\M$ is the
following.  If $C_1$ and $C_2$ are two loops in $Z$ that meet at a point
$p\in Z$, and $C_1*C_2$ is the loop made by ``joining'' $C_1$ and $C_2$
at $p$, one wishes
\eqn\nofc{H(C_1*C_2)=H(C_1)H(C_2).}
In the present case, this is proved as follows.  Let $D$ be a ``pair
of pants'' with three boundaries that we associate with $C_1$, $C_2$,
and $-C_1*C_2$.  (The minus sign refers to a reversal of orientation.)
Then the desired relation
\eqn\jofc{e^{iI(A_{C_1*C_2})}=e^{iI(A_{C_1})}e^{iI(A_{C_2})}}
follows from the definition \omore\ applied to the four-manifold
$X=\Sigma\times D$ whose boundary is the union of $X\times C_1$,
$X\times C_2$, and $-X\times C_1*C_2$.

One can verify directly
from the definition \hofc\ that if $C$ is a straight line
on the Jacobian from the origin to any point  $a\in H^1(\Sigma,\Z)$ (which
we represent by a harmonic one-form of the same name), then
$H(C)=1$.  This is done by first finding an oriented three-manifold
$B$, of boundary $\Sigma$, over which $a$ extends as a closed but no
longer harmonic one-form with integral periods.
($B$ can have very simple topology; it can be a ``handlebody.'')
Over the four-manifold $X=B\times \S^1$, the gauge field $A_C$ extends
in a fairly obvious way (as a $\theta$-dependent multiple of $a$, with
again $\theta$ parametrizing the position on $\S^1$) so
that $F\wedge F=0$ pointwise, and therefore $I_X(A)=0$.  
So as promised several times, 
 $\M$ is our friend, the modular-invariant
line bundle with all $H(a)=1$.

\nref\ramadas{T. R. Ramadas, I. M. Singer, and J. Weitsman,
``Some Comments On Chern-Simons Theory,'' Commun. Math. Phys.
{\bf 126} (1989) 409.} 
\nref\axelrod{S. Axelrod, {\it Geometric Quantization Of Chern-Simons
Theory}, Ph.D. thesis, Princeton University (1991).}
\nref\freed{D. Freed, ``Classical Chern-Simons Theory, Part 1,'' 
Adv. Math. {\bf 113} (1995) 237.}
The foregoing constructs an isomorphism class of line bundles $\M$ but
not an actual $\M$.  If one is interested in ``seeing'' an actual $\M$ as
precisely as possible, one may use the following somewhat abstract
approach (for full details and a number
of variants see \refs{\ramadas - \freed}).
Given a $U(1)$ gauge field $A$ on $\Sigma$, we must construct a one-dimensional
complex vector space $\M_A$, in such a way that the $\M_A$ vary nicely with 
$A$,
as fibers of a complex line bundle $\M$.  We simply declare that if
$N$ is any three-manifold with boundary $\Sigma$ over which $A$ extends
- -- that is if $N$ has boundary $(\Sigma,A)$ 
\foot{If $A$ is a connection on a topologically non-trivial line
bundle $\T$ over $\Sigma$, then such an $N$ does not exist.  Instead one fixes
a Riemann surface $\Sigma_0$ with a line bundle $\T_0$ of the same
first Chern class as $\T\to \Sigma$, and takes $N$ to be a bordism
between $\Sigma$ and $\Sigma_0$ (extending the line bundles and connection).
The rest of the construction proceeds as in the text.} --
then $\M_A$ has a basis vector $\psi_A(N)$.  Given two possible $N$'s, 
say $N_1$ and $N_2$, we must now exhibit a linear relation between
$\psi_A(N_1) $ and $\psi_A(N_2)$.  This linear relation is chosen
to be as follows.  Let $P$    be the oriented three-manifold obtained
by gluing $N_1$ and $N_2$ together (with opposite orientation for $N_2$)
along their common boundary.  Let $A_P$ be the connection (on a line bundle 
over $P$) made by combining the chosen extensions of $A$ over $N_1$ and $N_2$.
The relation between $\psi_A(N_1)$ and $\psi_A(N_2)$ is then chosen to be
\eqn\ubbu{\psi_A(N_1)=e^{iI(A_P)}\psi_A(N_2).}
Given many $N_i$'s, one would  construct
three-manifolds $P_{ij}$ by gluing $N_i$ to $N_j$, and set
$\psi_A(N_i)=e^{iI(A_{P_{ij}})} \psi_A(N_j)$; these  relations can be
shown to be compatible using the definition \omore\ of the
Chern-Simons functional.
Armed with these relations, the $\psi_A(N_i)$ can be interpreted
as vectors in a common one-dimensional space $\T_A$, which is the desired 
fiber of $\T$ over the connection $A$.  For a description of the connection
on $\T$ and explanation of its properties, consult the references.

\bigskip\noindent
{\it Chern-Simons Theory At Level One Half}

Now it is clear what we need in order to get a line bundle whose
first Chern class will be $\omega$ instead of $2\omega$.  We must
consider Chern-Simons theory at level one-half.

In other words, consider the functional
\eqn\oggo{{I(A)\over 2}= {1\over 16\pi}\int_X\epsilon^{ijkl}F_{ij}F_{kl},}
where $A$ is a gauge field on an oriented three-manifold $M$, and
$X$ is an oriented four-manifold, of boundary $M$,
 over which $A$ has been extended.
If $I(A)/2$ were well-defined (independent of the choice of $X$) modulo $2\pi$,
then by using everywhere $e^{iI(A)/2}$ instead of $e^{iI(A)}$ in the
above construction, we would
get the desired line bundle of first Chern class $\omega$.

The problem is that, for a closed four-manifold $Y$, the integral
\eqn\noggo{{1\over 16\pi}\int_Y\epsilon^{ijkl}F_{ij}F_{kl}}
is an arbitrary integer multiple of $\pi$, but not necessarily
an {\it even} integer multiple of $\pi$.  This comes from the following.
Let $x$ be the cohomology class $[F/2\pi]$.  Then the integral
in \noggo\ can be interpreted as $\pi x^2$.  In general,  the class
$x$ is integral, so $x^2$ is integral, but subject to no other
general restrictions.

Suppose, however, that $Y$ is a spin manifold.  If so,  the intersection
form on $H^2(Y,\Z)$ is even: for arbitrary $x\in H^2(Y,\Z)$,
$x^2$ is an {\it even} integer.  This is just what we need.
If we may assume
that the four-manifolds $Y$ in \noggo\ are always spin, then \noggo\ would
indeed always be an integral multiple of $2\pi$.

If one wants to encounter only four-manifolds with spin structure,
one must begin by only considering three-manifolds with a chosen spin
structure.  Thus, let $M$ be an oriented three-manifold with a chosen
spin structure.  Given a $U(1)$ gauge field $A$ over $M$, we would
like to define $I(A)/2$.  
(The definition will in general depend on the spin structure of $M$.)
The definition is made by the  formula
\oggo, where $X$ is an arbitrary oriented four-manifold of boundary $M$,
to which $A$ and the spin structure of $M$ extend.  (A proof that such
$X$'s exist was sketched in a previous footnote.)  Any two such
$X$'s will glue together to make a {\it spin}-manifold $Y$, for which
\noggo\ will be an integral multiple of $2\pi$.  Therefore, if $I(A)/2$ is
defined using only $X$'s of the indicated type, then  the definition of 
$I(A)/2$ in \oggo\ is independent of the choice of $X$ modulo $2\pi$.

Now let us tidy up a few details and solve our  problem.
We are given a Riemann surface $\Sigma$ with a spin structure $\alpha$.
We want to find a line bundle $\L_\alpha$ over the Jacobian 
$J_\Sigma$, with a connection,
compatible with the polarization of $J_\Sigma$.  Given a loop $C$
in the space of connections on $\Sigma$, we build (as above) the associated
gauge field $A_C=A_i(x;\theta)$ on the three-manifold $M=\Sigma\times \S^1$.
We give $M$ the spin structure which is the product of the spin structure
$\alpha$ on $\Sigma$ with the ``Neveu-Schwarz'' spin structure on $\S^1$
(this is the ``antiperiodic'' spin structure, the one which arises
if $\S^1$ is regarded as the boundary of a disc).  We then characterize
$\L$ by declaring  its holonomy  around $C$ to be
\eqn\onno{H(C)=e^{iI(A_C)/2}.}
The need to use 
the antiperiodic spin structure on $\S^1$ emerges  when one tries to prove
that $H(C_1)H(C_2)=H(C_1*C_2)$.  The proof of this involves the
four-manifold $\Sigma\times D$, where $D$ is a ``pair of pants'' (with
$C_1$, $C_2$, and $-C_1*C_2$ on the boundary). 
The reason that one must use the antiperiodic spin structure on the boundary
components is that (i) one needs to use the same spin structure on each
boundary component, to treat them all symmetrically (or the $H(C)$'s will
not obey the appropriate factorization); (ii) the spin structure on the
boundary of $D$ must extend over $D$.  The two conditions together
are satisfied precisely if the spin structure on the boundary is antiperiodic.

\bigskip\noindent{\it A Note On Bordism}

Happily, this is the end of the story for the chiral boson, at least
for the present paper.  But with an eye to generalizations, 
we will say a word about how bordism really enters the construction.

The above recipe for defining the line bundles $\L_\alpha$ used
the fact that if $M$ is a spin three-manifold with line bundle $\T$,
then there is a four-manifold $X$, with boundary $M$, over which the
spin structure of $M$ and the line bundle $\T$ extend.  This
is described mathematically by saying that $\Omega_3^{spin}(\CP^\infty)=0$,
a fact discussed in a previous footnote.

What would we say instead if $\Omega_3^{spin}(\CP^\infty)$ were non-zero?
To make the discussion definite, suppose that $\Omega_3^{spin}(\CP^\infty)$
were $\Z_2$.  Pick a three-manifold $M_0$ with a $U(1)$ gauge field $A_0$
such that the pair $(M_0,A_0)$ represents the non-zero element of
$\Omega_3^{spin}(\CP^\infty)$.  There is then no natural way to define
$I(A_0)/2$, since no appropriate $X$ exists for the definition \oggo.
However, if $\Omega_3^{spin}(\CP^\infty)=\Z_2$, then there exists
an $X$ whose boundary is {\it two} copies of $(M_0,A_0)$.  By
using this $X$ in \oggo, we can define $w=2\cdot I(A_0)/2$.
There are now two candidates (namely $w/2$ and $\pi+w/2$) for
what $I(A_0)/2$ should be.  Pick one of them.

Once $I(A_0)/2$ has been defined, one can define $I(A)/2$ for any
$U(1)$ gauge field $A$ on a three-manifold $M$.  In fact, the pair
$(M,A)$ is either the boundary of some $X$ or is bordant to
$(M_0,A_0)$ via some $X$ (that is, there is an $X$ whose boundary
is either $(M,A)$ or $(M,A)-(M_0,A_0)$, where the minus sign represents
a reversal of orientation).  Taking the integral in \oggo\ to define
either $I(A)/2$ or $I(A)/2 - I(A_0)/2$ as the case may be, we get
(since $I(A_0)/2$ has already been defined) a general definition of
$I(A)/2$.  This definition can be used to define the line bundle
$\L_\alpha\to J_\Sigma$ for every spin surface $(\Sigma,\alpha)$.
The definition is invariant under $\alpha$-preserving diffeomorphisms
of $\Sigma$, but will depend on the choice that was made in defining
$I(A_0)/2$.  Thus, we get a definition of the $\L_\alpha$'s that is
diffeomorphism-invariant to the expected extent, but depends on
a $\Sigma$-independent choice in the definition of $I(A_0)/2$.  This
choice is roughly analogous to the choice of a discrete theta angle.

The generalization from $\Z_2$ to an arbitrary group is easy to state.
Let $H={\rm Hom}(\Omega_3^{spin}(\CP^\infty),U(1))$.  Then $H$ classifies
the possible definitions of $I(A)/2$, and thus the possible
recipes for the association $(\Sigma,\alpha)\to \L_\alpha$.  The sort
of ``discrete theta angle'' that has just been described will turn
out not to appear for five-branes -- as the relevant bordism group
is zero -- but may appear in other, similar examples.

\newsec{Chiral Two-Form On Five-Brane World-Volume}

In this section, we will use a differential form notation to avoid
a proliferation of unilluminating constants.  For instance, if $M$
is a three-manifold with $U(1)$ gauge field $A$, and
$X$ is a four-manifold with boundary $(M,A)$, we write
$x=F/2\pi$, and we define the   Chern-Simons functional
simply as
\eqn\ohho{I(A)=2\pi\int_X x\wedge x.}
This is aimed to make it clear that -- as $\int_X x\wedge x$ would
be an integer for $X$ a {\it closed} four-manifold -- $I(A)$
is well-defined with values in $\R/2\pi \Z$.  We also will omit
wedge products and write $x^2$ for $x\wedge x$.

Now, consider a chiral two-form $\beta$ on a six-manifold $W$.
First we consider $W$ in isolation; then we consider a more general
case -- relevant to $M$-theory -- in which $W$ is embedded in an 
eleven-dimensional spin manifold $Q$, satisfying certain physical
conditions.

We know from section two that to study the chiral two-form,
we should introduce the intermediate Jacobian $J_W$ of $W$.  
This carries a natural polarization $\omega\in H^2(J_W,\Z)$, and
we must find a line bundle on $J_W$ whose first Chern class is $\omega$.
The partition function of the chiral two-form is then uniquely
determined.

We also know that to find a line bundle on $J_W$, we should
consider Chern-Simons theory in {\it seven} dimensions.  Let then
$M$ be a seven-manifold with a three-form field $C$.  
Actually, the field $C$ we want
 is not a three-form in the standard sense.  The field strength
$G=dC$ is allowed to have $2\pi$ periods (this statement will
soon receive a slight modification), and the gauge transformations
$C\to C+d\epsilon$ ($\epsilon$ a two-form) are to be supplemented
by ``big'' gauge transformations adding to $C$ a closed three-form
with $2\pi$ periods.  The relation of $C$ to a conventional three-form
is just like the relation of a $U(1)$ gauge field to a conventional 
one-form.  Anyway, given such a $C$, we want to define the Chern-Simons
functional $I(C)$.  This is done in a fashion that should be
familiar.  Let $X$ be an oriented eight-manifold with boundary
$M$ over which $C$ extends.\foot{Existence of such an $X$
depends on whether the pair $(M,G)$ vanishes in $\Omega_7(K(\Z,4))$
(which classifies  bordism classes of seven-manifolds with a 
four-dimensional cohomology class).  I do not know if that group
vanishes; if not, the considerations raised at the end of section
two will enter.  But presently we will put a spin condition on
$M$, and then the relevant bordism group becomes
$\Omega_7^{spin}(K(\Z,4))$, which does vanish by a result of
Stong \ref\otherstong{R. Stong, ``Calculation Of $\Omega_{11}^{spin}
(K(\Z.4))$,'' in {\it Unified String Theories}, eds. M. B. Green
and D. J. Gross (World Scientific, 1986).}.}  We describe this
in brief by saying that $X$ has boundary $(M,C)$.  Let $x=G/2\pi$.
The Chern-Simons functional is then
\eqn\norko{I(C) =2\pi\int_X x^2.}

This definition makes it clear that $I(C)$ is well-defined
modulo $2\pi$.  Therefore, $I(C)$ can be used exactly as in section
two to define a line bundle $\M$ over the intermediate Jacobian
$J_W$ of a six-manifold.  However, since the construction is 
completely diffeomorphism-invariant, $\M$ can hardly be the line
bundle we want.  In fact, rather as for the chiral scalar 
in two dimensions,
$\M$ is the line bundle
of first Chern class $2\omega$ and holonomy one around every straight line
in $J_W$.

To make progress, we must define $I(C)/2$ modulo $2\pi$.  Then
the Chern-Simons construction of line bundles, 
starting with $I(C)/2$, will
give the right line bundle.

To define $I(C)/2$ modulo $2\pi$, some additional structure is needed.
Recalling the case of the chiral scalar, and with an eye on physics,
in which spinors are present, one's first thought is to assume
that $M$ is a spin manifold with a chosen spin structure 
(and restrict $X$ so that the spin
structure of $M$ extends over $X$ -- such an $X$ exists as noted
in the last footnote).  If it were the case   that the intersection
form on $H^4(X,\Z)$, for an eight-dimensional spin manifold $X$, were
always even, our problem would be solved.  The even-ness of $x^2$ in
the spin case would give a factor of     two in \norko, so that $I(C)/2$
would be well-defined modulo   $2\pi$.  

It is, however, not true that the intersection form on the middle-dimensional
cohomology of an eight-dimensional spin manifold is even.  Indeed,
the quaternionic projective space ${\bf HP}^2$ is a simple
counterexample. ($H^4({\bf HP}^2,\Z)=\Z$, and the intersection
form is the unimodular, but not even, form with $f(x)=x^2$.) 
Instead there is the following relation.  Let $p_1$ be the  first
Pontryagin class of $X$.  In any dimension, there is a canonical
way to divide the first Pontryagin class of a spin manifold by two
to get an integral class that we will call $\lambda$.  Then
for any $v\in H^4(X,\Z)$,
\eqn\jubbo{v^2=v\cdot \lambda ~~~~{\rm modulo}\,\,2.}
(A proof of this using $E_8$ index theory is at the end of section four
of \witten.)  This can be rewritten as the statement that
\eqn\onormo{{1\over 2}\int_X\left(
\left(v-{\lambda\over 2}\right)^2-{\lambda^2\over 4}\right)\in \Z.}

This suggests that we should be the following.  Instead of asking
that $G/2\pi$ should be integral, we require
\eqn\ikko{\left[G\over 2\pi\right]={\lambda\over 2}- v,}
with $v$ an integral class.  This is in fact the correct
quantization law for $G$ in $M$-theory \witten.  Then, with $x$ still 
denoting $G/2\pi$, we modify the definition of $I(C)$ slightly and take
\eqn\ubbu{{\tilde I(C)\over 2}=\pi\int_X\left(x^2-{\lambda^2\over 4}\right).}
Since $x=\lambda/2-v$ with integral $v$, it follows from \onormo\
that $\tilde I(C)/2$ is well-defined modulo $2\pi$ and so can be
used to define a line bundle $\L$ (just as we would have done with
$I(C)/2 $ had that been well-defined modulo $2\pi$
on the original class of $G$'s).

As we have divided by two, $\L$ has the desired first Chern class,
$c_1(\L)=\omega$.
We do have to be a little careful in describing where $\L$ is defined.
Because the quantization law of $G$ has been shifted, $G=0$
may not be allowed.  To be more precise, \ikko\ permits $G=0$
precisely if $\lambda$ is even.
If $\lambda$ is not even, the above construction gives a line bundle
$\L$ not on the intermediate Jacobian $J_W$ but on a shifted version
of it with $G/2\pi$ congruent to $\lambda/2$ modulo integral classes.

In the application to $M$-theory, with $W$ understood as the world-volume
of a five-brane, one can assume (because of the world-volume equation
$dT=G$) that the restriction of $G$ to $W$ has vanishing cohomology class. 
(And \ikko\ means therefore that the restriction of $\lambda$ to $W$ must
be even.)  In this situation, it is the ordinary, unshifted
intermediate Jacobian $J_W$ on which the above construction gives
a line bundle.  

The main point is that as  the correction to quantization of
$G$ needed to define $\tilde I(C)/2$ is the same as the one that
appears in the physics,  the line bundle appears precisely in the right
place.

\subsec{Embedding In Eleven Dimensions}

So far, we have considered the chiral two-form on a ``bare''
six-dimensional spin manifold $W$.

We are really interested in an $M$-theory application in which
$W$ is a six-dimensional submanifold of an eleven-manifold $Q$.
Moreover, $W$, though oriented,
is not necessarily spin; it is $Q$ that    carries a spin structure.

We will assume for simplicity that $W$ is compact.  More general
cases, in which for instance $W$ is asymptotically flat, are also
natural; in such  cases, some knowledge of what is happening
at infinity can serve as a substitute for compactness.  

The discussion in the last subsection was adequate if the normal
bundle to $W$ in $Q$ is trivial -- that is if $Q$ looks locally near
$W$ like $W\times \R^5$.  If so, the normal directions can be
decoupled from the discussion and what we are about to say reduces
to what was said above.

The first question we might want to ask is how to achieve
gauge-invariance even locally
for the three-form field $C$ of $M$-theory.  
We recall that $C$ is coupled to the chiral two-form $\beta$
on the five-brane, and that this coupling is not invariant
under gauge transformations $\delta C=-d\alpha$ of the $C$-field.
The failure of gauge invariance was described in \onorko:\foot{
Of course, \onorko\ is the violation of gauge invariance for
a non-chiral two-form with chiral coupling to $C$, while we
want a chiral two-form.  The difference between the two is crucial
in discussing subtle global issues such as those considered in 
this paper, but not for studying local perturbative anomalies.}
\eqn\hobby{\delta L_{ eff}\sim \int_Q \alpha\wedge G.}
The numerical coefficient multiplying the right hand side can be
most usefully described as follows.  Suppose that $W$ is the boundary
of a seven-manifold $M$. Let $I(C)$  be the seven-dimensional
Chern-Simons functional defined in \norko.  It is gauge-invariant
on a closed seven-manifold, but not on a seven-manifold with boundary.
In fact,
\eqn\obby{\delta L_{ eff}={1\over 2}\delta I(C).}
This factor of one-half is the reason that in the discussion above
it was necessary to define a version of $I(C)/2$.  Upon  taking
$W=\Sigma \times {\bf CP}^2$, with $\Sigma$ a Riemann surface much
larger than the ${\bf CP}^2$, the chiral two-form on $W$ reduces
to a chiral scalar on $\Sigma$, and the factor of $1/2$ in \obby\
reduces to the factor of one-half difference between \duggo\ and \bimo\
which was essentially the subject of section two.

We must find another interaction in the theory that can cancel the anomaly
\obby.
In the first instance, this interaction is simply the classical
interaction $I_0={\rm const}\cdot 
\int_QC\wedge G\wedge G$ of eleven-dimensional supergravity.
The normalization of $I_0$ can be most usefully described by noting
that if $Q$ is the boundary of a twelve-manifold $Y$ over which $G$ 
extends, and $x=G/2\pi$, then
\eqn\ombii{I_0=-{2\pi\over 6}\int_Y x\wedge x \wedge x.}
The factor of $-1/6$, which was important in \witten, is related
to $E_8$ index theory as explained there.

To prove that $I_0$ is  invariant under $\delta C=-d\alpha$,
one must integrate by parts and use $dG=0$.  But in the field
of a five-brane, $dG$ is not zero; it is a delta function supported
on      the five-brane world-volume $W$.  So instead of being
zero, the gauge variation of $I_0$ is a multiple of $\int_W \alpha\wedge
G$, and can potentially cancel \hobby.
That the cancellation actually occurs is a consequence of the
formula $(1/2)+3(-1/6)=0$, where the 
$1/2$ is present in \obby, the $-1/6$ in \ombii, and the 3 reflects
the fact that in comparing the eleven-dimensional interaction
$I_0\sim C GG$ to the seven-dimensional interaction $I_1\sim CG$,
one takes one of the three fields in $I_0$ to be in the normal
direction to $W$ and two to be tangential; there are three ways to do 
this.

What we have just analyzed is a piece of perturbative anomaly
cancellation for five-branes, namely the term involving $C$ only.
One should also consider perturbative gravitational anomalies
for five-branes.  Some such terms were studied
in \refs{\duff,\fivewitten}; a more complete discussion is in section
five below.  Because of invariance under sign change of $C$ together with
reversal of orientation of the normal bundle, there are no ``mixed''
$C$-gravitational perturbative anomalies.

\bigskip\noindent{\it Definition Of The Line Bundle}

We now have the crucial clue for how to find the desired
line bundle
$\L\to J_W$ and therefore the desired partition function.
In the previous examples (chiral boson and chiral two-form on
a ``bare'' six-manifold), the key was to find a Chern-Simons
interaction (in a higher dimension) that is gauge-invariant
on a closed manifold but not in the presence of a boundary.
We now do exactly the same thing, but the higher dimension
will be  eleven (and not seven, as the previous experience might
suggest), and the word ``boundary'' must be replaced by ``five-brane.''

Thus, in $M$-theory, there is a Chern-Simons interaction, schematically
 $I_M =CGG +CI_8(R)$, where the $CGG$ term  
is the interaction $I_0$ considered above, and $I_8(R)$ is a certain
quartic polynomial in the Riemann tensor \refs{\vw,\duff}.
The expression
\eqn\pinoo{  W=\det\, D_{R.S.} \,\,e^{iI_M},}
with $\det\,D_{R.S.}$ the Rarita-Schwinger path integral,
is well-defined \witten\
on a closed eleven-manifold $Q$.\foot{The bordism statement used
here is of course Stong's theorem \otherstong\ that 
$\Omega_{11}^{spin}(K(\Z,4))=0$.}  
The two factors are not separately well-defined, but as explained
in \witten, one can alternatively factor \pinoo\ as follows:
\eqn\hinoo{W=\left\{\det\, D_{R.S.}\,e^{iI_{R.S.}/2}\right\}
\cdot e^{iI_{E_8}}.}
Here $I_{R.S.}$ is a properly normalized Chern-Simons term
related to the Rarita-Schwinger operator, and $I_{E_8}$ is
a properly normalized Chern-Simons term related to $E_8$ index theory.
The virtue of the factorization in \hinoo\ is that the $C$-dependence
has been put entirely in $J=e^{iI_{E_8}}$, which is a conventional
Chern-Simons term, so our general framework will apply.  
The factor in curly brackets in \hinoo\ will
play practically no role in the discussion.

On an eleven-manifold
$Q$ that has a boundary $R$, $J$ is not gauge-invariant in the
usual sense, but rather must be interpreted as a section of
a line bundle, which we will now call $\L^{-1}$, over the space of
fields on $R$.  This line bundle $\L^{-1}$ can be described somewhat
more concretely by arguments given in section two.  Those
arguments, after all, had a purely formal character: given
{\it any} Chern-Simons interaction -- such as $I_{E_8}$ -- which
is well-defined on a closed manifold but not on a manifold with
boundary,
 one always produces a line bundle over  a suitable space
of fields on the boundary.
Once $\L^{-1}$ is found, the chiral two-form partition function
must be a section of the inverse line bundle $\L$ (to cancel global
as well as local anomalies), and therefore its partition function
is determined.

Let us now go over some of that ground in a little more fully.
Five-branes may sound different from boundaries, but for the present 
purposes, the two are quite similar.  If $W$ is a five-brane
world-volume in a space-time $Q$, then the $G$ field has a singularity
along $W$.  Since singularities are awkward at best and the low
energy field theory description is really not valid near the singularity,
one might want to cut out of space-time a small tubular neighborhood
of $W$.  The boundary of that neighborhood is a ten-manifold 
$R$ which is an $\S^4$ bundle over $W$. The fact that $W$ was a five-brane
world-volume is now captured by saying that is $S$ is a fiber
of $R\to W$, then 
\eqn\uccuu{\int_R{G\over 2\pi}=1.}

A sensible definition of the chiral two-form partition function
should depend only on the local geometry near $W$ -- which
means, apart from a knowledge of $W$ itself, only a knowledge
of the normal bundle to $W$ in $Q$ and a choice of spin structure
on a neighborhood of $W$ in $Q$.  All this information can be summarized
by giving the ten-manifold $R$, together with the map $R\to W$
obeying \uccuu, and a spin structure on $R$.  (As the normal bundle
to $R$ in $Q$ is a trivial real line bundle, the spin structure on
$Q$ induces one on $R$.)  To obtain a definition of the chiral
two-form partition function that only depends on the geometry
near $Q$, we should show that  we can make the definition given
only $R$.

The first step is to define a line bundle $\L^{-1}$ over the
space of $C$-fields on $R$, with an action of the local and global
gauge transformations.  For this we need a well-defined Chern-Simons
action on a closed eleven-manifold; we choose for this the object
$I_{E_8}$.  With this Chern-Simons action, the construction
in section two now produces a line bundle $\L^{-1}$ over
the space of $C$-fields on $R$.  We recall that this is done
as follows: given a loop $L$ in the space of $C$ fields on $R$,
one builds a $C_L$ field on $R\times \S^1$, and declares the
holonomy of $\L^{-1}$ around $L$ to be the value of $e^{iI_{E_8}}$
on $R\times \S^1$ with $C=C_L$, the antiperiodic
spin structure on $\S^1$,  and the product metric on
$R\times \S^1$.\foot{An important detail must be
checked here.  The factorization in \hinoo, by which  we
eliminated the determinant and reduced to Chern-Simons theory,
is unique only up to $I_{E_8}\to I_{E_8}+I'$, where $I'$ is
a properly normalized Chern-Simons interaction constructed
from the metric only.  To really have a unique construction of
$\L^{-1}$, the holonomy around the loop $L$ should be independent
of the choice of factorization.  This is true since
 $R\times \S^1$ with
product metric is the boundary of $R\times D$ with product
metric ($D$ being a two-disc), and the relevant curvature polynomials
are all zero pointwise on $R\times D$ with product metric.}  

To finish, then, all we need is a map $i$ from $C$-fields on $W$ to
$C$-fields on $R$ that obey \uccuu\ 
(commuting with local and global gauge
transformations).   Given such a map, we use $i$  to ``pull back''
the line bundle $\L^{-1}$ from the space of $C$-fields on $R$ to
the space of $C$-fields on $W$, and this gives finally 
the gauge-invariant line bundle we need on the space of $C$-fields
on $W$.  

The desired $i$ is found as follows.  Let $\pi:R\to W$ be the
projection, and let $\tau:R\to R$ be the map that commutes with
$\pi$ and acts as $-1$ on each $\S^4$ fiber of $\pi$. A solution
$C_0$ of
\uccuu\  can be described uniquely up to gauge transformation
by saying that $G_0=dC_0$ is harmonic
(remember that $C_0$ can have Dirac string singularities, so that this
is possible) and that the field is odd under $\tau$.\foot{For
existence of such a field, take any solution $C_0$ 
of \uccuu\ with harmonic
$G_0$, and replace it by  $(C_0-\tau^*C_0)/2$.
For uniqueness, note that if $C_0$ and $C'_0$ both obey the
conditions, then $C'=C_0-C'_0$ is a closed three-form odd under
$\tau$ and therefore (since $\tau$ acts as $+1$ on $H^3(R,\Z)$,
which is isomorphic to $H^3(W,\Z)$) vanishes up to gauge transformation.} 
The desired map from $C$-fields on $W$ to $C$-fields on $R$ is
$C\to \pi^*C+C_0$.  

Via $i$, we pull back $\L^{-1}$ to the space of $C$-fields on
$W$.  Restricting to $C$-fields with $G=0$ and dividing by
local and global gauge transformations, we get a line bundle
$\L^{-1}$ over $J_W$.  The  chiral two-form partition function
is determined by the fact that it is a section of the dual line
bundle $\L$.  

\newsec{Chiral Four-Form In Ten Dimensions}

Now we move on to the other somewhat similar example: a
chiral four-form $\gamma$ in ten dimensions, which appears in Type IIB
superstring theory.

The field strength of $\gamma$ is a self-dual five-form $L$.
It does not obey $dL=0$, as one might have guessed.   Rather,
if $B^i, \,i=1,2$ are the two two-forms of
Type IIB supergravity (which of course transform in the two-dimensional
representation of $SO(2,\Z)$), and $H^i=dB^i$, then the relation is
\eqn\oppop{dL=\epsilon_{ij}H^i\wedge H^j.}
This means that if we set $E=\epsilon_{ij}B^i\wedge H^j$, then
$E$ behaves as a sort of ``composite five-form gauge field'' that
is coupled to $\gamma$, just as a $U(1)$ gauge field can be coupled
to a chiral scalar in two dimensions, and the $C$-field of
eleven-dimensional supergravity is coupled to the chiral
two-form on a five-brane world-volume.  

>From section two, we know that the partition function of the
chiral four-form $\gamma$ on a ten-dimensional spin manifold $W$
will be a section of a line bundle $\L$ over the intermediate
Jacobian $J_W$, and that finding the partition function is equivalent
to finding a line bundle $\L$ whose    first Chern class equals
the polarization $\omega$ of $J_W$.  We also know that
we can always use Chern-Simons theory to find a line bundle of
first Chern class $2\omega$, and that we can use Chern-Simons
theory to find a line bundle of first Chern class $\omega$ provided
that, for a closed twelve-dimensional spin manifold $X$, the intersection
form on $H^6(X,\Z)$ is always even.  
This last statement is, happily, true (unlike its counterpart in
eight dimensions, whose falsehood made the last section more complicated).

More generally, in fact, the intersection form on the middle
dimensional cohomology of a closed spin manifold is always even
in $8k+4$ dimensions.  A proof of this using the Adem relations
for the Steenrod algebra has been described by J. Morgan.
\foot{The argument is as follows.  In $8k+4$ dimensions, one
has a relation in the Steenrod algebra $Sq^{4k+2}=Sq^2Sq^{4k}
+Sq^1Sq^{4k}Sq^1$.  Let $x$ be an element of $H^{4k+2}(X,\Z_2)$,
with $X$ an $8k+4$ dimensional spin manifold.  Then modulo two,
one has $x^2=Sq^{4k+2}(x)=Sq^2(Sq^{4k}x)+Sq^1(Sq^{4k}Sq^1x)$.
The right hand side vanishes, since $Sq^2$ and $Sq^1$, as maps
to the top dimension, are multiplication by $w_2$ and $w_1$, which
vanish for a spin manifold.}

One point that should be made about the chiral four-form in
ten dimensions is that in using Chern-Simons theory to define
the line bundles, the bordism group that one meets is
$\Omega_{11}^{spin}(K(\Z,6))$. (This is the bordism group
of an eleven-manifold endowed with a six-dimensional
class; it enters in the same way that $\Omega_3^{spin}(K(\Z,2))
=\Omega_3^{spin}({\bf CP}^\infty)$ entered in section two.)
 As far as I know, this group
has not been computed, and might conceivably be non-zero.
If so, then as described at the end of section two, a sort
of exotic ``theta angle'' would appear in the theory, 
parametrized by the dual
group $H={\rm Hom}(\Omega_{11}^{spin}(K(\Z,6)),U(1))$.  Such
an extra parameter in Type IIB superstring theory seems 
unlikely, so one might conjecture that in fact $\Omega_{11}^{spin}
(K(\Z,6))=0$.

\bigskip\noindent{\it An Example}

It may be helpful to present a concrete example, to show
that if we are given a specific ten-manifold $W$ (or similarly
a specific
six-manifold and normal bundle in the five-brane case)
we actually can use these considerations to find the line  bundle
$\L$ and therefore the partition function of the chiral field.

The example I will consider is $ W=\S^5\times \S^5$.  Call the two
factors $\S^5_1$ and $\S^5_2$.  The
middle-dimensional cohomology is $H^5(W,\Z)=\Z\oplus \Z$, with
the two $\Z$'s coming cohomology of the two $\S^5$'s.  The intermediate
Jacobian is $J_W=\R^2/(\Z\oplus \Z)$.  
Let $a$ and $b$ be the lattice points $(1,0)$ and $(0,1)$ in
$\Z\oplus \Z$.  There are two distinguished
straight lines $C_a$ and $C_b$
in $J_W$, induced by the straight lines from
the origin to  $a$ or $b $.  I claim that the line bundle
$\L$ made from the Chern-Simons construction has 
holonomy 1 around $C_a$ or $C_b$.  According to \keycon,
this determines the holonomy around all cycles and
uniquely determines the line bundle $\L$.  (For instance, the holonomy
on a straight line from the origin to $(1,1)$ is $-1$.)

The Chern-Simons construction involves the coupling of the chiral
four-form to a background five-form $E$.  (In Type IIB superstring
theory, $E$ is a composite field, found in the discussion of 
\oppop, but for the present purposes $E$ might as well be elementary.)
For instance, to compute the holonomy around $C_b$, we must
consider the eleven-manifold $ M=\S^5_1\times \S^5_2\times \S^1$,
with an $  E$ -field which is the pullback of an $E$-field
on $\S^5_2\times \S^1$, and such that if $K=dE$ is the
field strength, then $\int_{\S^5\times\S^1}K=1$.
  ($E$ may also be taken to vary linearly in the $\S^1$ 
direction, but we will not need this.)  Now, we can regard $\S^5_1$
as the boundary of a ball $B^6$, over which the spin structure
extends.  So $M$ is the boundary of $X=B\times \S^5_2\times \S^1$.
The $E$-field on $X$ can be a pullback of an    $ E$-field on
$\S^5_2\times \S^1$.
The holonomy around $C_b$ is to be computed from
$\int_X K\wedge K$.  But this vanishes, since ($E$ being a
pullback from the last two factors), $K\wedge K$ actually vanishes
point-wise.

So the chiral four-form partition function on $\S^5\times \S^5$ 
is determined.  As a mathematical corollary, note that it follows
that,
since $\S^5\times \S^5$ has only one spin structure, which is
therefore preserved by all orientation-preserving diffeomorphisms,
 the diffeomorphism group of $\S^5\times \S^5$ does not
induce the full $SL(2,\Z)$ action on $H^5(\S^5\times \S^5,\Z)$, but
at most the subgroup preserving the particular line bundle $\L$.  (This
subgroup can be shown to be of index three.)
By contrast, in the superficially similar case of $\S^1\times
\S^1$, the diffeomorphism group does induce the full $SL(2,\Z)$ action
on $H^1(\S^1\times\S^1,\Z)$.  

\newsec{Perturbative Anomaly Cancellation For Five-Branes}

In this section, we re-examine perturbative anomaly cancellation
for $M$-theory five-branes.  We have already verified anomaly
cancellation for the $C$-field gauge invariance in section three.
It remains to consider gravitational anomalies.  It turns out
that this leads to a surprisingly long story -- nothing about
five-branes seems to be straightforward! -- and we will get a complete
answer only for Type IIA, not for $M$-theory.

The five-brane world-volume is a six-manifold $W$ in an eleven-manifold
$Q$.  $W$ is oriented, and (though this requirement can
be relaxed, as $M$-theory
conserves parity) we will consider only the case that $Q$ is oriented.
Let $TQ$ be the tangent bundle to $Q$ and
 $TQ|_W$ its restriction to $W$.  We have  $TQ|_W=TW\oplus N$,
where $TW$ is the tangent bundle to $W$ and $N$ is the normal bundle to
$W$ in $Q$.  $N$, in particular, is an $SO(5)$ bundle over $W$.

Note that  a Riemannian metric on $Q$ induces a Riemannian
metric on $W$, and a metric and $SO(5)$ connection on $N$.
The theory along $W$ therefore has some features of gravity
coupled to $SO(5)$ gauge theory.

In analyzing anomalies,
it is enough to consider only diffeomorphisms of $Q$ that map $W$ to $W$,
since the presence of the five-brane wrapped over $W$ explicitly
breaks other diffeomorphisms.  A diffeomorphism of $Q$ that maps $W$ to
$W$ induces first of all a diffeomorphism of $W$ and secondly 
an $SO(5)$ gauge transformation of the normal bundle.  In fact,
diffeomorphisms of $W$ and $SO(5)$ gauge transformations are all that
the the world-volume fields (and anomalous interactions) ``see'' (at
least in the long wavelength limit that suffices for computing
anomalies) 
so the discussion will amount to an analysis of gravitational
and $SO(5)$ gauge anomalies on $W$.

Perturbative anomalies in $2n$ dimensions are always related to
characteristic classes in $2n+2$ dimensions, so in the present case
- -- as $W$ 
is six-dimensional -- the anomalies will involve eight-dimensional
characteristic classes.  It turns out to be rather helpful to write
the anomalies in terms of Pontryagin classes $p_i(TW)$ and $p_i(N)$.
The anomaly eight-form is then {\it a priori} a linear combination of 
$p_2(TW)$,
$p_1(TW)^2$, $p_1(N)p_1(TW)$, $p_1(N)^2$, and $p_2(N)$.
The terms involving $p_i(TW)$ only  have been analyzed before
\refs{\duff,\fivewitten}; we will extend the analysis to include the
other terms.  As we will see, the discussion is surprisingly 
unstraightforward.  The known and expected contributions to
the $p_1(N)p_1(TW)$ and $p_1(N)^2$ anomalies will cancel, but something
new is involved in the cancellation of the $p_2(N)$ term.

There are three known sources of anomalies: (1) world-volume fermions;
(2) the chiral two-form; (3) the Chern-Simons couplings of the
bulk theory, whose gauge invariance fails in the presence of the
five-brane.
Their contributions can be determined as follows.

\bigskip\noindent
{\it  World-Volume Fermions}
\def\ch{{\rm ch}}

The world-volume fermions are (four-component)
chiral spinors on $W$ with values in the
(rank four) bundle  $S(N)$ constructed from
$N$ by  using the spinor representation
of $SO(5)$.    According to standard anomaly formulas \ref\alv{L.
Alvarez-Gaum\'e and E. Witten, ``Gravitational Anomalies,''
Nucl. Phys. {\bf B234} (1983) 269.}, the contribution
of these fields to the anomaly is
\eqn\ollo{I_D={1\over 2}\ch S(N)\cdot \hat A(TW),}
where the $1/2$ arises because of considering chiral spinors,
 $\ch $ is the Chern character, 
and up through dimension eight
\eqn\hollo{\hat 
A(TW)=1-{p_1(TW)\over 24}+{7p_1(TW)^2-4p_2(TW)\over 5760}.}
To compute $\ch S(N)$, we note that if the Chern roots of the $SO(5)$
bundle $N$ are $\pm \lambda_1$, $\pm \lambda_2$, and 0, then the
Chern roots of $ S(N)$ are $\pm (\lambda_1\pm \lambda_2)/2$.
So up through terms quartic in the $\lambda$'s,
\eqn\opoppo{\eqalign{
\ch S(N)&=\sum_{\epsilon_1,\epsilon_2=\pm 1}
\exp\left[{{1\over 2}(\epsilon_1\lambda_1+\epsilon_2\lambda_2)}\right]
\cr
&= 4+{1\over 2}(\lambda_1^2+\lambda_2^2)+{1\over 96}(\lambda_1^4+\lambda_2^4
+6\lambda_1^2\lambda_2^2)\cr
&= 4+{p_1(N)\over 2}+{p_1(N)^2\over 96}+{p_2(N)\over 24}.\cr}}
In the last step we used $p_1(N)=\lambda_1^2+\lambda_2^2$,
$p_2(N)=\lambda_1^2\lambda_2^2$.  The detailed form of $I_D$
can be obtained by combining the last three equations.

\bigskip\noindent
{\it The Chiral Two-Form}

The chiral two-form propagates on $W$ and does not ``see'' the normal
bundle.  The standard anomaly of such a field is
\eqn\umbo{I_A={1\over 5760}\left(16p_1(TW)^2-112p_2(TW)\right).}

\bigskip\noindent
{\it Anomaly Inflow From The Bulk}

The third term to consider is the anomaly inflow from the bulk.
The eleven-dimensional bulk theory has an interaction proportional to
$\tilde I=C\wedge (p_1(TQ)^2/4-p_2(TQ))$, where $p_1(TQ)$ and $p_2(TQ)$ can be
understood as certain polynomials in the Riemann tensor of $Q$.  It is
important that what appears
in $\tilde I$ is the Riemann tensor (and therefore the Pontryagin classes)
of $Q$, not those of $W$; after all, $\tilde I$ is an interaction defined
in the eleven-dimensional bulk theory.  The relevance to perturbative
anomalies of 
$\tilde I$ is that, although gauge-invariant in bulk, it is not gauge-invariant
in the field of a five-brane.  The anomaly $I_B$ coming from this
term is
\eqn\huuh{I_B=-{1\over 48}\left({p_1(TQ|_W)^2\over 4}-p_2(TQ|_W)\right).}
As $TQ|_W=TW\oplus N$, 
we have $p_1(TQ|_W)=p_1(TW)+p_1(N)$, $p_2(TQ|_W)=p_2(TW)+p_2(N)
+p_1(N)p_1(TW)$.  \huuh\ can thereby be rewritten
\eqn\turtle{I_B=-{1\over 48}\left({p_1(TW)^2+
p_1(N)^2-2p_1(TW)p_1(N)\over 4} 
- -p_2(TW)-p_2(N)\right).}

Upon summing up \ollo, \umbo, and \turtle, one finds that all terms
involving $p_i(TW)$ cancel (both the purely gravitational term and
the ``mixed'' term $p_1(TW)p_1(N)$), as does the $p_1(N)^2$ term.
The remaining anomaly is in
fact
\eqn\indigo{\delta = {p_2(N)\over 24}.}

So something new is needed.  It turns out that it is easier to understand
the new ingredient  if one considers the problem in Type IIA
superstring theory rather than in $M$-theory.  
In other words, we take $Q$ to be $M\times \S^1$, where $M$ is 
a ten-manifold, and we consider the case that the five-brane world-volume
$W$ is a submanifold of $M$ (times a point in $\S^1$).  
$M$-theory then becomes equivalent to Type IIA superstring theory,
and the five-brane becomes the solitonic five-brane of Type IIA, which
couples magnetically to the string theory
two-form $B$ (of field strength  $H=dB$). 

The normal bundle $N$ then becomes $N=N'\oplus {\cal O}$ where ${\cal O}$ 
is a trivial one-dimensional bundle (representing the tangent bundle
to $\S^1$) and $N'$ (the tangent bundle to $W$ in $M$) is an $SO(4)$ 
bundle.  In particular, $p_i(N)=p_i(N')$ for $i=1,2$.

One thing that  is special about $SO(4)$, however, is that for an $SO(4)$
bundle -- such as $N'$ -- $p_2$ can be written in terms of a four-dimensional
class, called the Euler class of the bundle, $\chi(N')$.
This has its roots in the fact that at the Lie algebra level, $SO(4)=
SU(2)\times SU(2)$, so that an $SO(4)$ bundle has two four-dimensional
characteristic classes -- $p_1$ and $\chi$ -- related to the instanton
numbers in the two $SU(2)$'s.  These are the independent characteristic
classes of an $SO(4)$ bundle
(in general a Lie group of rank $r$ has $r$ independent
such characteristic classes). So in particular
$p_2$ can be written in terms of these.
The relation is in fact $p_2(N')=\chi(N')^2$.   $p_2$ can be
derived from the function $\lambda_1^2\lambda_2^2$ of the Chern roots
and $\chi$ can be derived from the function $\lambda_1\lambda_2$.
An expression $p_2=\chi^2$ can exist in $SO(4)$ but not in $SO(5)$
because
the function $\lambda_1\lambda_2$ is Weyl-invariant in $SO(4)$ -- whose
Weyl group acts on the  $\lambda_i$ with pairwise sign changes
 -- but not in $SO(5)$, whose Weyl group generates
independent sign changes of the $\lambda_i$. 

Some notation concerning the Euler class will be helpful.  We represent an
object $\alpha$ transforming in
the adjoint representation of $SO(4)$ by a $4\times 4$ antisymmetric
tensor $\alpha^{ij}$, $i,j=1,\dots,4$.  In particular, the curvature
$F$ of an $SO(4)$ connection $A$ is a two-form $F^{ij}$ with values
in that representation.  The Euler class is represented by the four-form
$\chi(F)=\epsilon_{ijkl}F^{ij}\wedge F^{kl}/32\pi^2$, 
where $\epsilon_{ijkl}$ is the
fourth rank invariant antisymmetric tensor of $SO(4)$.  Locally, given
a choice of gauge, $\chi(F)$ is the exterior derivative of a Chern-Simons
three-form that we call $\Omega_\chi(A)$: $\chi(F)=d\Omega_\chi(A)$.  
If $\alpha$ is an $SO(4)$ gauge generator, we write $\chi(\alpha,F)
=\epsilon_{ijkl}\alpha^{ij}F^{kl}/16\pi^2$.  
The gauge variation $\delta_\alpha\Omega_\chi$ of $\Omega_\chi$
under a gauge transformation by $\alpha$ is 
\eqn\ujju{\delta_\alpha\Omega_\chi(A) = d\left(\chi(\alpha,F)\right).}
This equation is part of the ``descent'' formalism familiar
in the study of anomalies.

\bigskip\noindent
{\it The Euler Class}

The anomaly of interest is thus -- in the Type IIA context -- $\chi(N')^2/24$,
and it will be necessary to have some understanding of the particular
meaning of the characteristic class $\chi(N')$.

The basic question we have to focus on is: what is the $H$-field 
(that is, the three-form field strength of Type IIA) produced
by a five-brane world-volume $W$ in a ten-manifold $M$?  The following
considerations only involve the behavior near $W$, and so only depend
on the topology of  the normal bundle $N'$ to $W$ in $M$.  We can
in fact replace $M$ with the total space of $N'$.

The $H$-field is supposed to be a three-form such that $dH=\delta_W$,
where $\delta_W$ is a delta function supported on $W$.
Such an $H$ is not uniquely
determined, as one could add any smooth, closed $H$-field.
However, there is an obstruction to existence of $H$: the obstruction
is that $\chi(N')$ must vanish.  This is explained in \ref\bott{R. Bott
and L. W. Tu, {\it Differential Forms In Algebraic Topology}
(Springer-Verlag, 1982).}, beginning on p. 70.

The essence of the problem is to precisely formulate what $\delta_W$ is
supposed to be.  $\delta_W$ should be a four-form on $M$ which is
closed, is supported in a small neighborhood of $W$ in $M$, and
(if one identifies $M$ with the total space of $N'$)
integrates to 1 over any fiber of $N'\to W$.  

If $B$ is a differential form on a manifold $M$, the restriction of
$B $ to a submanifold $W$, written $B|_W$, is obtained by considering
only the values on $W$ of components of $B$ tangent to $W$.  Thus
$B|_W$ is a differential form on $W$ (of the same degree as $B$).
The theory of the Thom isomorphism and the Euler class, as described in
\bott, shows that any $\delta_W$ with the properties stated in the last
paragraph has the
further property that $\delta_W|_W$ is in the cohomology class
of $\chi(N')$, the Euler class of the normal bundle $N'$.  If
a connection $A$ on $N'$ is picked, so that $\chi(N')$ is represented
by a differential form $\chi(F)$ (as defined above), then one can
in a very natural way pick $\delta_W$ such that
\eqn\omy{\delta_W|_W=\chi(F).}

Now it is clear why there is a restriction on the possible topology of
the normal bundle to a five-brane.
The ``magnetic'' field of the five-brane is supposed to be a three-form
$H$ obeying
\eqn\gomy{dH= \delta_W,}
but this implies 
\eqn\tomy{d(H|_W)=\delta_W|_W=\chi(F).} 
Such an $H$  can exist only if the differential form
$\chi(F)$ is trivial cohomologically, that is only if the characteristic
class $\chi(N')$ vanishes.  

A form with the properties
of $\delta_W$ could be constructed using {\it any}
connection $A$ on $N'$, but when $W$ is embedded in $M$ as a five-brane
world-volume, there is a distinguished connection -- coming from
the Riemannian connection of $M$ -- and therefore a distinguished
four-form $\chi(F)$ representing the Euler class of the normal bundle. 
\gomy\ is part of the naive idea of
what a five-brane wrapped on $W$ is supposed to be, but is not
usually stated precisely enough to exhibit the ``finite part'' of
the delta function along $W$ that appears in \tomy.  
We will assume in the rest of this
paper that \tomy\ should be taken as part of the    definition of
a five-brane.

Since the anomaly we are 
trying to eliminate is proportional to $\chi(N')^2$,
the vanishing of $\chi(N')$ means that the integrated anomaly vanishes
in a suitable sense.  That is not enough; we need to cancel the anomaly
locally, since gauge transformations are local.  But vanishing of the
integrated anomaly at least means that there is no topological obstruction
to finding a counterterm that would cancel the anomaly.  In fact
it is easy to see, using \tomy, that there is such a counterterm.
It is
\eqn\jummy{\tilde L= \int_W H|_W\wedge \Omega_\chi(A),}
where again $A$ is the Riemannian 
connection on $N'$ and $\Omega_\chi$  was introduced above.
Indeed, using \ujju\ and \tomy\ and integrating by parts, 
the variation of $\Delta L$ under gauge transformations of $N'$ is 
\eqn\bummy{\delta_\alpha \tilde L =-\int_W\chi(F)\wedge \chi(\alpha,F),}
and this is the six-form related by the usual ``descent'' procedure
to the anomaly eight-form $\chi(F)^2/2$.

So a multiple of $\tilde L$ will cancel the $\chi(N')^2$ perturbative
anomaly.  Moreover, given the invariance of Type IIA superstrings under
reversal of orientation of $N'$ together with sign change of $H$,
this is the only anomaly that can be canceled by such a term.
It is thus gratifying that precisely this term is the one whose
 contributions from previously known interactions do not cancel.

\bigskip\noindent
{\it Back To $M$-Theory}  

For $M$-theory five-branes, we need a generalization of this,
but how to do so is somewhat puzzling.  
The replacement of $SO(4)$ by $SO(5)$ and of $\chi(N')^2$ by
$p_2(N)$ makes even the absence of a topological obstruction
to canceling the anomaly mysterious.  For a specific five-brane
world-volume $W$, $p_2(N)$ vanishes, as $W$ is six-dimensional.
But the topological interpretation of anomalies involves
considering certain two-parameter families of physical objects
\ref\as{M. F. Atiyah and I. M. Singer, ``Dirac Operators Coupled
To Vector Potentials,'' Proc. Natl. Acad. Sci. US {\bf 81} (1984) 2597.},
 and one could perfectly
well have a two-parameter family of $W$'s with non-zero $p_2(N)$
over the total space.  It is not clear what sort of physical
mechanism would suppress such families in $M$-theory.

Beyond canceling the topological obstruction, we need to actually
find a local counterterm that cancels the $p_2(N)$ anomaly.  The
fact that this counterterm must reduce to \bummy\ under the
appropriate conditions makes it clear roughly what the desired
counterterm must be, though the matter still seems rather obscure.
For an $SO(5)$ bundle $N$, the closest analog
of the characteristic class $\chi$ is 
a certain four-form with values in $N$, defined by
\eqn\ombimbo{\chi_e={F^{ab}\wedge F^{cd}\epsilon_{abcde}\over 32\pi^2}.}
The factorization $p_2=\chi^2$ that holds for $SO(4)$ bundles
becomes for $SO(5)$ bundles
\eqn\nobimbo{p_2(N)=\sum_e\chi_e\wedge \chi_e.}
In $M$-theory, the fundamental differential form in space-time is
not a three-form $H$ but a four-form $G$.  However, along $W$, one
can define an $N$-valued three-form, namely the part of $G$ with
three indices tangent to $W$ and one $N$-valued index.  Let us write this
part of $G$ as $H_e$, where $e$ is the $N$-valued index, and
the   indices tangent to $W$ are not written explicitly.
The analog of \jummy\ must be something like
\eqn\uxxx{\int_W \sum_eH_e\wedge \Omega_e}
where $\Omega_e$ is an $N$-valued three-form related to $\chi_e$.
$H_e$ must also obey an $N$-valued version of \tomy. 
It is not clear exactly what the right equations are.  The fact
that the story works so nicely for Type IIA nevertheless gives
some faith that a satisfactory answer must exist in $M$-theory.
\listrefs  
\end